\documentclass[10pt,conference]{IEEEtran}

\usepackage{amsmath}
\usepackage{multirow}
\usepackage{booktabs}
\usepackage{array}
\usepackage{float}
\usepackage{tikz}
\usepackage{latexsym}
\usepackage{enumerate}
\usepackage{colortbl}
\usepackage{threeparttable}
\usepackage{subfigure}
\usepackage{graphicx}
\usepackage{cite}

\newcommand{\tabincell}[2]{\begin{tabular}{@{}#1@{}}#2\end{tabular}}

\usetikzlibrary{arrows,decorations.markings}
\usetikzlibrary{calc,trees,positioning,arrows,chains,shapes.geometric,%
	decorations.pathreplacing,decorations.pathmorphing,shapes,%
	matrix,shapes.symbols}

\begin{document}

\title{Connecting Software Metrics across Versions to Predict Defects}

\IEEEoverridecommandlockouts
\author{
	Yibin Liu$^{1,2,*}$,  Yanhui Li$^{1,2,*}$, Jianbo Guo$^{3,*}$, Yuming Zhou$^{1,2}$, Baowen Xu$^{1,2}$ \\
	
	\small
	1. State Key Laboratory for Novel Software Technology, Nanjing University, China \\
	\small 2. Department of Computer Science and Technology, Nanjing University, China\\
	\small 3. Institute for Interdisciplinary Information Sciences, Tsinghua University, Beijing, China\\
	\thanks{*These authors contributed equally to this work.}
}

%*Equal contribution. $\dagger$Corresponding author.

\maketitle
\begin{abstract}

Accurate software defect prediction could help software practitioners allocate test resources to defect-prone modules effectively and efficiently. In the last decades, much effort has been devoted to build accurate defect prediction models, including developing quality defect predictors and modeling techniques. However, current widely used defect predictors such as code metrics and process metrics could not well describe how software modules change over the project evolution, which we believe is important for defect prediction. In order to deal with this problem, in this paper, we propose to use the Historical Version Sequence of Metrics (HVSM) in continuous software versions as defect predictors. Furthermore, we leverage Recurrent Neural Network (RNN), a popular modeling technique, to take HVSM as the input to build software prediction models. The experimental results show that, in most cases, the proposed HVSM-based RNN model has a significantly better effort-aware ranking effectiveness than the commonly used baseline models. 
%In particular, the proposed HVSM-based RNN model can identify many real defects missed by the commonly used baseline models. 
\end{abstract} 

%\keywords{Defect prediction, historical version sequence, code metrics, recurrent neural network, classification techniques}

\section{Introduction}
%Fixing software defects is a very important part of software maintenance that consumes a huge amount of time and effort \cite{Erlikh2000Leveraging}. The important preliminary step of bug fixing is to identify which part contains bugs.  An extensive body of work \cite{techniques.1,background.dp1,techniques.3,background.dp2,techniques.2,SementicFeatures,dp.Unsupervised,Yang2016} has been proposed in software defect prediction that can detect the defect-prone parts of the software and help software engineers test and debug. The main purpose of defect prediction is to build  prediction models to tell whether a new software part is buggy or not. 
Fixing software defects is a very important part of software maintenance that consumes a huge amount of time and effort \cite{Erlikh2000Leveraging}. The preliminary step of bug fixing is to identify the potential locations of bugs in a software project. In the last decades, many software defect prediction models \cite{techniques.1,background.dp1,techniques.3,background.dp2,techniques.2,SementicFeatures,dp.Unsupervised,Yang2016} have been proposed to identify defect-prone modules, which could help software engineers test and debug software more effectively and efficiently. In order to achieve accurate defect prediction, it is essential to use quality defect predictors and modeling techniques to build the prediction models. 

There are two main aspects of concern in the establishment of prediction models: one is proposing new views to collect metrics or selecting proper metrics to improve the prediction performance; the other is introducing new classification techniques that can perform better and comparing them with previously used techniques.  

In terms of metrics for defect prediction, most existing studies concentrate on two types of metrics: code metrics and process metrics. The code metrics can well describe the static characteristics of a file in a given version, and a proper classifier could group those similar files together to distinguish buggy files from clean ones.  CK features \cite{metrics.CK} and McCabe features \cite{metrics.McCabe} are the commonly studied static metrics which are extracted from code static properties (e.g., dependencies, function and inheritance counts). Recently, Wang et al. \cite{SementicFeatures} leveraged Deep Belief Network (DBN) to automatically learn semantic features which is a more complex kind of code metrics to capture the semantic difference of source code. Process metrics have been provided to describe the change information in project's development. Bird et al. \cite{Bird2011} examined the effect of ownership in defect prediction and provided process metrics that measure the contribution of developers in projects. Mockus and Weiss \cite{Mockus2000} studied the performance of metrics measuring the change of a project in predicting the risk of new changes. Arisholm et al. \cite{Arisholm2010A} and Rahman and Devanbu \cite{Rahman2013} both investigated the predictive power of different types of metrics including code and process metrics, and drew the same conclusion that using process metrics can significantly improve the performance of prediction. McIntosh et al. \cite{McIntosh2016} also studied the predictive power of code review coverage, participation and expertise, and found them effective in predicting defects. In most cases, process metrics are version-duration, which means that they are extracted from two adjacent versions to describe the change of files between the two versions.
%Moser et al \cite{Moser2008A} suspected that change data would be more helpful to predict defect proneness of source files than source code itself. They compared code metrics and change metrics for predicting files of the Eclipse project and found that process metrics are more efficient than code metrics.

Meanwhile, researchers have applied many machine learning classification algorithms to build prediction models in software defect prediction \cite{background.dp5,Rparameter}. Wang and Li . \cite{Wang2010Naive} constructed a Naive Bayes (NB) based model to predict software defects on 
MDP datasets, which has better performance than decision tree based learner J48. Sun et al. \cite{Sun2012Using} compared the performance of several types of classification algorithms over the 14 NASA datasets, and found that Random Forest (RF) is not
significantly better than the others. Zhang et al. \cite{dp.Unsupervised} proposed a connectivity-based unsupervised classifier, Spectral Clustering, and compared the performance of SC with supervised classifiers using data from 26 projects. 

\begin{figure}
	\centering
	\includegraphics[width=0.5\textwidth]{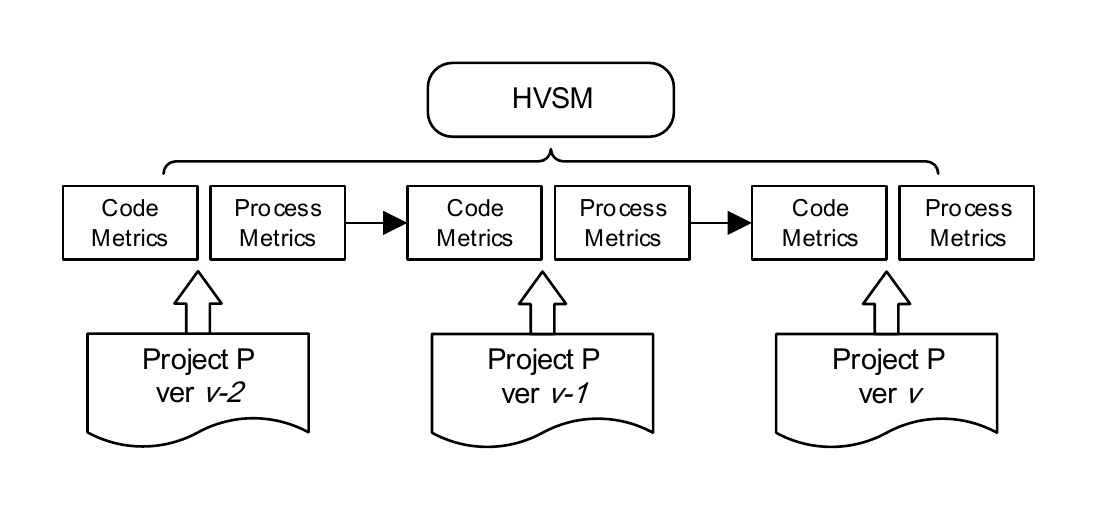}
	\caption{An overview of different types of metrics}
	\label{intro}
\end{figure}

This paper deviates from existing studies in two important ways. \textbf{Firstly,} we propose a new way to construct predictors based on software metrics: we connect a file's metrics in several continuous versions together in ascending order of version (Figure \ref{intro}). We call this new predictor \textbf{Historical Version Sequence of Metrics} (HVSM). Compared with code metrics and process metrics, HVSM has the following advantages: 
\begin{itemize}
	\item HVSM provides a new and more complete perspective for engineers and managers to consider and explain the trend of how the files change over the project's evolution. 
	\item HVSM employs code metrics to describe files' static data or process metrics to describe change data. Additionally, the sequence of HVSM can reveal the files' comprehensive change history which are not included in most process metrics.
	%\item HVSM reuses the existing code metrics of the project, and joins them at a relatively low cost. It is convenient for the researchers to construct HVSM from code metrics available on open data repository, such as PROMISE. 
\end{itemize}
\textbf{Secondly}, we bring in a new classification technique, Recurrent Neural Network (RNN)  \cite{rnnreview}, to process the HVSMs. Since a long lived file may exist in more versions, and has a longer HVSM than others, the HVSM set of the project may contains variable-length HVSMs. The existing techniques \cite{Elish2008Predicting,Sun2012Using,Wang2010Naive,dp.Unsupervised} of defect prediction cannot directly handle variable-length metrics in HVSMs. An RNN is a powerful graphical model for sequential data inputs, which uses the internal state to memorize previous inputs and handle variable length inputs. 

Our approach consists of three steps: a) constructing the historical version sequence of each file in our projects; b) extracting the HVSM of each file from its version sequence; (c) leveraging RNN to predict file-level defects using the extracted HVSMs (details are in Figure \ref{figure.HVSM} and Section \ref{section.approach}). 

The main contributions of this paper are listed as follows:
\begin{itemize}
	\item We provide a novel view to reveal the static and change information of a file in the version sequence of a project, which is expressed in the form of HVSM in defect prediction.
	\item We leverage HVSM with the help of a classification technique RNN to improve the performance in within-project defect prediction (WPDP) and evaluate the result of our approach compared with 7 typical classifiers on 9 open source projects from \textsc{Promise}. The SK test and Win/Tie/Loss evaluation shows that our approach outperforms other techniques in effort-aware scenarios. 
	%The overlap result also shows a high complementarity between our approach and the baseline classifiers.
\end{itemize} 

The rest of this paper is organized as follows. Section \ref{section.background} describes the background and related work on defect prediction and RNN. Section \ref{section.approach} introduces our approach to extract HVSMs and applies RNN to perform defect prediction. Section \ref{section.experiment} shows the experimental setup. Section \ref{section.result} evaluates the performance of our approach against other techniques. Some discussions and threats to validity of our work are presented in section \ref{section.discussion} and \ref{section.threats}. We conclude our work in section \ref{section.conclusion}.

\section{Background and Related Work}
\label{section.background} 
This section provides the background and related work of file-level defect prediction and recurrent neural network.

\subsection{Typical Defect Prediction Process}
Defect prediction on file-level has been studied by many prior works \cite{background.dp5,background.dp1,background.dp2,background.dp3,related.CLAMI,background.wpdp,background.dp4,background.dp6,SementicFeatures,background.dp7,data.2}. The typical process of file-level defect prediction is as follows: firstly, labeling data as buggy or clean based on bug reports for each file; secondly, collecting corresponding metrics of these files as features; thirdly, training prediction models using machine learning classifiers with the input of instances with features and bug labels; Finally, predicting whether a new instance is buggy or clean using trained models. 

There are two main scenarios of defect prediction, one of them is within-project defect prediction (WPDP) \cite{background.wpdp2,background.dp2,multiMetrics.MWY,f1.threshold,background.wpdp,SementicFeatures}. In WPDP, researchers always train classifiers using the data in an older version and predict defects in a newer version within the same project. The other one is cross-project defect prediction (CPDP) \cite{background.cpdp,background.dp4,SementicFeatures,Hosseini:2016:SBT:2972958.2972964}. The CPDP problem is motivated by the fact that many companies lack the training data in practice. A typical solution for CPDP is to apply prediction models that are built using data from a different source project. 

%and cross-validation defect prediction (CVDP) \cite{multiMetrics.MWY,background.wpdp2,Prateek2013,Rparameter}. 
%The common practice is to implement cross-validation in a specific version of a project \cite{multiMetrics.MWY,background.wpdp2}. 
In this study, we focus on improving the performance in WPDP with our approaches.

\subsection{Recurrent Neural Network} 
\label{section.RNN}

%classicrnn
\begin{figure}
	\centering
	\begin{tikzpicture}[shorten >=2pt, auto, node distance=\layersep,thick]
	
	\tikzstyle{every pin edge}=[<-,shorten <=1pt]
	\tikzstyle{neuron}=[circle,fill=black!25,minimum size=18pt,inner sep=0pt]
	\tikzstyle{hidden neuron}=[neuron, draw=black];
	\tikzstyle{output neuron}=[neuron, draw=black];
	\tikzstyle{input neuron}=[neuron, draw=black];
	\tikzstyle{input metric}=[neuron, fill=white];
	\tikzstyle{annot} = [text width=8em, text centered];
	\tikzstyle{vecArrow} = [thick, decoration={markings,mark=at position
		1 with {\arrow[semithick]{open triangle 60}}},
	double distance=1.4pt, shorten >= 5.5pt,
	preaction = {decorate},
	postaction = {draw,line width=1.4pt, white,shorten >= 4.5pt}]
	\newcommand{\first}{-5}
	\newcommand{\unfolding}{-1.5}
	\newcommand{\forward}{2.5}
	\node[annot] (Model) at (\first,1.5) {Hidden Layer};
	\node[annot] (Model) at (\first,0) {Input Layer};
	\node[annot] (Model) at (\first,3) {Output Layer};
	\node[input neuron] (I) at (\unfolding,0) {$x$};
	\node[hidden neuron] (H) at (\unfolding,1.5) {$S$};
	\path (H) edge [loop left,->,] node{$W$} (H);
	\node[output neuron] (O) at (\unfolding,3) {$P$};
	\path (I) edge[->] node{$U$} (H);
	\path (H) edge[->]  node{$V$} (O);
	\end{tikzpicture}
	\caption{A RNN model with one hidden layer. }
	\label{classicrnn}
\end{figure}
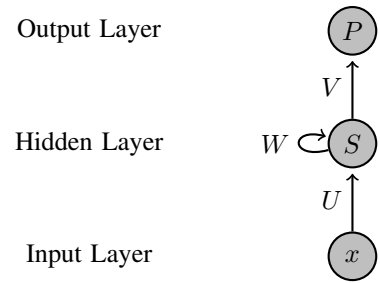

%figure.HVSM
\begin{figure*}
	\center
	\resizebox{1\textwidth}{!}{
		\begin{tikzpicture}
		\tikzstyle{vecArrow} = [thick, decoration={markings,mark=at position
			1 with {\arrow[semithick]{open triangle 60}}},
		double distance=1.4pt, shorten >= 5.5pt,
		preaction = {decorate},
		postaction = {draw,line width=1.4pt, white,shorten >= 4.5pt}]
		
		%\foreach \x in {1,2.5,4,5.5,7}
		%\foreach \y in {1,2,3,4}
		%\draw (\x ,\y) grid +(1,1);
		
		\node[text width=3em, text centered] at (0, 5.5) {Version:};
		\node[text width=3em, text centered] at (0, 4.5) {Module:};

		\foreach \x/\y in {1/3,2.5/2,4/1}
		\node[text width=3em, text centered] at (\x + 0.5, 5.5) {$ v-\y$};
		
		\node[text width=3em, text centered] at (5.5 + 0.5, 5.5) {$v$};
		
		\node[text width=3em, text centered] at (7 + 0.5, 5.5) {$v+1$};
		
		% \foreach \x in {1,3,5,7}
		% \node[text width=3em, text centered]  at (\x + 0.5, 0.5){$\vdots$};
		
		\node (spade-1) at (1.5,4.5) {$\spadesuit$};
		\node (spade-2) at (3,4.5) {$\spadesuit$};
		\node (spade-3) at (4.5,3.5) {$\spadesuit$};
		\node (spade-4) at (6,4.5) {$\spadesuit$};
		\node (spade-5) at (7.5,4.5) {$\spadesuit$};
		
		\node at (7.5,3.5) {$\cdots$};
		\node at (7.5,2.5) {$\cdots$};
		\node at (7.5,1.5) {$\cdots$};
		
		\foreach \x/\y in {1/2,2/3,3/4,4/5}
		\draw[blue,thick] (spade-\x) --  (spade-\y);

		\node (heart-1) at (3,2.5) {$\heartsuit$};
		\node (heart-2) at (4.5,2.5) {$\heartsuit$};
		\node (heart-3) at (6,3.5) {$\heartsuit$};
		
		\foreach \x/\y in {1/2,2/3}
		\draw[blue,thick] (heart-\x) --  (heart-\y);
		
		\node (diamond-1) at (4.5,1.5) {$\diamondsuit$};
		\node (diamond-2) at (6,2.5) {$\diamondsuit$};
		
		\foreach \x/\y in {1/2}
		\draw[blue,thick] (diamond-\x) --  (diamond-\y);
		
		\node (club-1) at (6,1.5) {$\clubsuit$};
		
		\node (box-1) at (1.5,3.5) {$\Box$};
		\node (box-2) at (3,3.5) {$\Box$};
		\node (box-3) at (4.5,4.5) {$\Box$};
		
		\foreach \x/\y in {1/2,2/3}
		\draw[blue,thick] (box-\x) --  (box-\y);
		
		\node (triangle-1) at (1.5,2.5) {$\triangle$};
		\node (triangle-2) at (3,1.5) {$\triangle$};
		
		\foreach \x/\y in {1/2}
		\draw[blue,thick] (triangle-\x) --  (triangle-\y);
		
		\node (nabla-1) at (1.5,1.5) {$\nabla$};
		
		% \draw[->] (8.1,3) to (9.9,3);
		%\node[text width=4em,text centered] at (9,3.2) { HVSM$^v$};
		
		\foreach \x/\y/\z in {1/1/-3,2.5/2/-2,4/3/-1,5.5/4/ }
		\node (spaden-\y) at (\x +11,4.5) {$M_\spadesuit^{v\z}$};
		
		\foreach \x/\y in {1/2,2/3,3/4}
		\draw[blue,thick,->] (spaden-\x) --  (spaden-\y);
		
		\foreach \x/\y/\z in {2.5/2/-2,4/3/-1,5.5/4/}
		\node (heartn-\y) at (\x +11,3.5) {$M_\heartsuit^{v\z}$};
		
		\foreach \x/\y in {2/3,3/4}
		\draw[blue,thick,->] (heartn-\x) --  (heartn-\y);
		
		\foreach \x/\y/\z in {4/3/-1,5.5/4/}
		\node (diamondn-\y) at (\x +11,2.5) {$M_\diamondsuit^{v\z}$};
		
		\foreach \x/\y in {3/4}
		\draw[blue,thick,->] (diamondn-\x) --  (diamondn-\y);
		
		\node (clubn-1) at (5.5+11,1.5) {$M_\clubsuit^v$};
		
		\foreach \y/\x in {1/\clubsuit,2/\diamondsuit,3/\heartsuit,4/\spadesuit}
		\node at (10.7, \y +0.5) {HVSM$_\x^v$:};

		%\node at (10.7,5.5) {HVSM$^v$};
		
		\draw[decorate,decoration={brace}] let
		\p1=(spade-3.east), \p2=(spade-1.west) in  (6.5,1.2) -- (1,1.2)  node{};
		
		\node at (3.5,0.7) {HVSM$^{v}$ Extraction};
		
		% \draw (8,6) to (8.1,3);
		
		\draw[->,thick] (5.5,0.7) -- (9,0.7) -- (9,3) -- (9.7,3);
		
		% \draw (5.5,6) to (8,6);
		
		\draw[decorate,decoration={brace}] let
		\p1=(spade-1.east), \p2=(spade-4.west) in (1,5.8) -- (8,5.8) node{};
		
		\node at (4.5,6.3) {HVSM$^{v+1}$ Extraction}; 
		
		\draw[->,thick] (6.3,6.3) -- (9.7,6.3); 
		
		\foreach \x/\y/\z in {1/1/-3,2.5/2/-2,4/3/-1,5.5/4/,7/5/+1 }
		\node (spaden2-\y) at (\x +11,6.3) {$M_\spadesuit^{v\z}$};
		
		\foreach \x/\y in {1/2,2/3,3/4,4/5}
		\draw[blue,thick,->] (spaden2-\x) --  (spaden2-\y); 
		
		\node at (10.7,6.3) {HVSM$_\spadesuit^{v+1}$:};
		
		\node at (11+8.5, 7) {\textbf{Labels}};
		
		%\node at (13.5, 7) {\large \textbf{HVSM}}; 
		
		%\node at (4.5, 7) {\large \textbf{Version Sequence}};
		
		\node at (11+8.5, 6.3) {?};
		
		\node at (11+8.5, 1.5) {$\surd$};
		
		\node at (11+8.5, 2.5) {$\times$};
		
		\node at (11+8.5, 4.5) {$\surd$};
		
		\node at (11+8.5, 3.5) {$\times$};
		
		\draw[decorate,decoration={brace}] let
		\p1=(club-1.south), \p2=(spade-4.north) in (19.8,5) -- (19.8,1) node{};
		
		\node(vi) at (19.9, 3) {};
		
		\node[rectangle,draw=black,text width=4em,text centered,fill=black!25,minimum size=40pt,](comparision) at (23,6.3) {RNN Classifier};
		
		\draw [->,thick] (vi) to [bend right=45]  (comparision);
		
		\node at (21.5, 4) {\textbf{Training}};
		
		\draw [<-,thick] (19.9,6.3) -- (comparision);
		
		\node[text width=8em] at (21.5,6.25) {\textbf{Prediction:}\\ Buggy or not};
		
		\node at (4.5, 0) {\large \textbf{(a) Historical Version Sequence}};
		
		\node at (13.5, 0) {\large \textbf{(b) Historical Version Sequence of Metrics}}; 
		
		\node at (21.5,0 ) {\large \textbf{(c) RNN}};

		%  \node at (10.7,0.5) {$\vdots$};
		
		\end{tikzpicture}
	}
	\caption{Overview of our proposed HVSM-based defect prediction}
	\label{figure.HVSM}
\end{figure*}
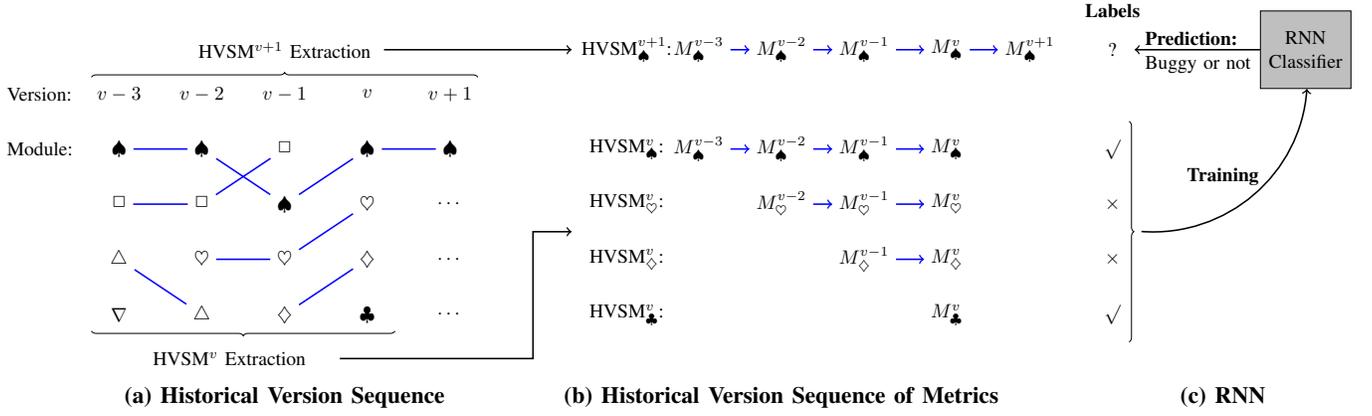

A recurrent neural network is a powerful graphical model for sequential data, which can handle variable length inputs or outputs for different demands, like speech recognition, video analysis and  natural language translation \cite{speech1,speech2,rnnreview}. Recently, RNN has been applied in related software engineering problems, especially in obtaining an API usage sequence for a given natural language query based on RNN Encoder-Decoder \cite{rnn.api}.

Figure \ref{classicrnn} shows the main structure of RNN, which contains one input layer, several self-connected hidden layers and one output layer. RNN processes information from a sequential input  $x^1, x^2, ..., x^T$, where $T$ is the length of the sequential data, by incorporating it into the hidden state $S$ ($S^1, S^2, ...,S^T$) that is passed through time, and outputs $P=P^T$ in the end which is combined with the corresponding training target $y$ to get the loss $L$ to train the network. $U$, $V$ and $W$ are the weight matrix between input and hidden layer, the weight matrix between hidden and output layer, and the weight matrix between hidden layer and itself at adjacent time steps respectively. The hidden state $S^t$ at every time step $t$ of the sequence is modeled as follows:
\begin{equation}\label{hiddenstate} 
S^t=
\begin{cases}
tanh(Ux^t+b), if ~ t=1\\
tanh(Ux^t+WS^{t-1}+b), if ~ t \in \{2,3, ..., T\}\\
\end{cases}
\end{equation}
where $tanh(z)=\frac{e^{z}-e^{-z}}{e^{z}+e^{-z}}$ and the vector $b$ are the bias parameters. Current activations $S^t$ in the hidden layer are determined by both current input $x^t$ and previous hidden layer activations $S^{t-1}$, which makes the network's internal state \textbf{memorize} previous inputs. The memory power of RNN is the  core reason we select RNN to deal with sequential inputs. And the output, i.e. the probability for $y=1$,  is:
\begin{equation}
P^T=sigmoid(VS^T+c)
\end{equation}
where $sigmoid(c)=\frac{1}{1+e^{-c}}$ and $c$ is the bias parameter. Once $P^T$ is obtained, the loss $L$ is computed as:
\begin{equation}
L =-ylogP^T-(1-y)log(1-P^T)+\frac{\lambda}{2}||\omega||_2^2 \\
\end{equation}
where $\frac{\lambda}{2}||\omega||_2^2$ is the so called Tikhonov regularization to avoid over-fitting on the training set:
\begin{equation*}
||\omega ||_2^2 =\sum_{i,j}U_{ij}^2+\sum_{i,j}V_{ij}^2+\sum_{i,j}W_{ij}^2
\end{equation*} 

RNN can be trained to minimize the loss $L$ using gradient descent with a technique known as back-propagation \cite{backprop1, backprop2}. We first initialize $U$, $V$ and $W$ randomly, then set $b$ and $c$ to $\bf{0}$, and update them by the iteration process:
\begin{equation}\label{gd} 
\begin{cases}
U_{ij}(l+1)&=U_{ij}(l)-\eta \frac{\partial L}{\partial U_{ij}}\\
V_{ij}(l+1)&=V_{ij}(l)-\eta \frac{\partial L}{\partial V_{ij}}\\
W_{ij}(l+1)&=W_{ij}(l)-\eta \frac{\partial L}{\partial W_{ij}}\\
b_k(l+1)&=b_k(l)-\eta \frac{\partial L}{\partial b_k}\\
c(l+1)&=c(l)-\eta \frac{\partial L}{\partial c}
\end{cases}
\end{equation}
where $l$ is the $l^{th}$ iteration, $\eta$ is the learning rate, $U_{ij}, V_{ij}, W_{ij}$ are the $i^{th}$ row, $j^{th}$ column entry of matrix $U, V, W$ respectively, $b_k$ is the $k^{th}$ entry of bias vector $b$. 

\section{Our Approach}
\label{section.approach}
In this work, we focus on the sequential information of files across versions. For files that exist in several continuous versions, the comprehensive sequential information in change history is useful when performing defect prediction. We define \textbf{Historical Version Sequence of Metrics (HVSM)} to highlight the version sequence information of files. With HVSM, a classification technique RNN is used to predict defective files. Our approach consists of three major steps: (1) constructing file's historical version sequence (2) extracting HVSMs from files' version sequences, (3) leveraging RNN to predict defects using the extracted HVSM. 

\subsection{Constructing Historical Version Sequence}
\label{section.HVS}
Considering the development of a project, files change across versions. As shown in Figure \ref{figure.HVSM}(a), each symbol represents a file in the project and they exist in different versions from $v-3$ to $v+1$. The connection lines between versions indicate the sequential change of the same file. We sort all versions containing a file $a$ in ascending order, and call it $a$'s historical version sequence. From Figure \ref{figure.HVSM}, taking version $v$ as the current version, the version sequence of file $\spadesuit$ lasts from $v-3$ until now, while file $\Box$ has its version sequence end at the previous version $v-1$. In our approach, we consider the files with the same name and the same path as one specific file when processing file's version sequence.

It should be noticed that, in common, files always exist in continuous versions. To summarize, we define 3 types of files at the time of a specific version $v$:
\begin{itemize}
\item \textbf{Developing File} ($\spadesuit$, $\heartsuit$, $\diamondsuit$): the files that are created in previous versions, and still exist in current version.
\item \textbf{Newborn File} ($\clubsuit$): The files that are created in current version.
\item \textbf{Dead File} ($\nabla$, $\Box$, $\triangle$): The files that exist in previous versions and disappear in current version.
\end{itemize}

The high percentage of developing files will make the historical information more complete when constructing historical version sequence. 

%change
For the project shown in Figure \ref{figure.HVSM}, assume that engineers want to predict defects in version $v+1$ using the versions before. A common practice \cite{SementicFeatures} is to use files in version $v$ to build the training set for a classifier. Code metrics and process metrics extracted from these files are frequently used in defect prediction. Considering the change history of files in a project, process metrics apparently provide more information than code metrics. Prior works extract process metrics from code ownership \cite{Bird2011}, change frequency\cite{Arisholm2010A,Rahman2013}, developer experience\cite{Mockus2000} and etc. Others also extract change metrics from version history \cite{Arisholm2010A} or use the metric difference between versions as features \cite{metric.delta}. 

The approaches above do take the change and process information of files into account, but most of the metrics are version-duration, which means that the metric considers only the information between two adjacent versions and cannot reveal the historical information of whole version sequence.

\subsection{Extracting HVSM}
\label{section.extracting}
Here we introduce the \textbf{Historical Version Sequence of Metrics (HVSM)} as mentioned at the beginning of this section to highlight the sequential information of file's changes across versions. 
%In practice, a file may exist in a very long version sequence, which makes it hard to extract the complete historical version sequence of it. In this case, the length of version sequence is limited when extracting HVSMs. 
For a specific length $len$, HVSM joins a file's metrics of at most $len$ continuous versions traced back from version $v$, and groups the metrics in ascending order of versions. For convenience, we introduce some symbols: for a given file $a$ in version $x$, we denote $M_a^x$ as its metric set; for $a$ in a studied version $v$, we define $HVSM_a^v$ as: 
\[HVSM_a^v=M_a^{v_o}\rightarrow M_a^{v_o+1}\rightarrow...\rightarrow M_a^v \]
where $v_o$ is the first version during the last $len$ versions, that $a$ exists. And we denote $T_a^v=v-v_o+1$ as the number of versions file $a$ exists from $v_o$ to $v$, and call it the length of $HVSM_a^v$.
For example in Figure \ref{figure.HVSM}, let $len=4$, the file $\heartsuit$ has $T_{\heartsuit}^v=3$. Assuming the metric set $M_{\heartsuit}^x$ containing 10 metrics, then its $HVSM_{\heartsuit}^v$ includes $10\times3=30$ metric values. It should be noticed that $1\leq T_a^v \leq len$. For a specific software project, we denote $HVSM^v$ as the set containing the HVSMs of all files in version $v$ of this project:
\begin{equation*}
HVSM^v=\{HVSM_a^v|a\in \text{all files in version } v\}
\end{equation*}
and $m^v=|HVSM^v|$ denotes the number of files in version $v$ of the project. Then we define the length of $HVSM^v$ as the number $len$ of concerned versions: $length(HVSM^v)=len$

Take the project in Figure \ref{figure.HVSM} as an example. 
For $len=4$, the $HVSM^v$ that contains HVSMs of the four listed files $\spadesuit$, $\heartsuit$, $\diamondsuit$ and $\clubsuit$ are shown in the right part of the figure. 
The dead files $\nabla$, $\Box$ and $\triangle$ do not exist in version $v$ so they are not included in $HVSM^v$. When predicting defects in $v+1$, the test set is also built with HVSM as the example $HVSM^{v+1}_\spadesuit$ shown in the figure.
%To be noticed that the performance of defect prediction when using HVSM with different $len$ may differ, we also conduct experiments to compare the predictive power of HVSM with different $len$ (see Section \ref{section.rq3}).

%In this paper, we focus on defect prediction, so if it is not specially mentioned, HVSM will refer to $HVSM^v$ or a specific part of $HVSM^v$ (like training set and test set in classification models). 

%considering only the developing files and the newborn files for a specific version. In the version $v$, the $HVSM^v$ that contains HVSMs of the four listed files $\spadesuit$, $\heartsuit$, $\diamondsuit$ and $\clubsuit$ are shown in the right part of the figure. The dead files $\nabla$, $\Box$ and $\triangle$ don't exist in version $v$, so their HVSMs are not included in $HVSM^v$. In typical techniques, files in each version are equivalent to samples for classifiers and independent across versions. With HVSM, the characteristics of code metrics and process metrics are considered and HVSM contains the file's complete information during its lifetime. The sequential information that HVSM implied will help improve the performance of prediction.

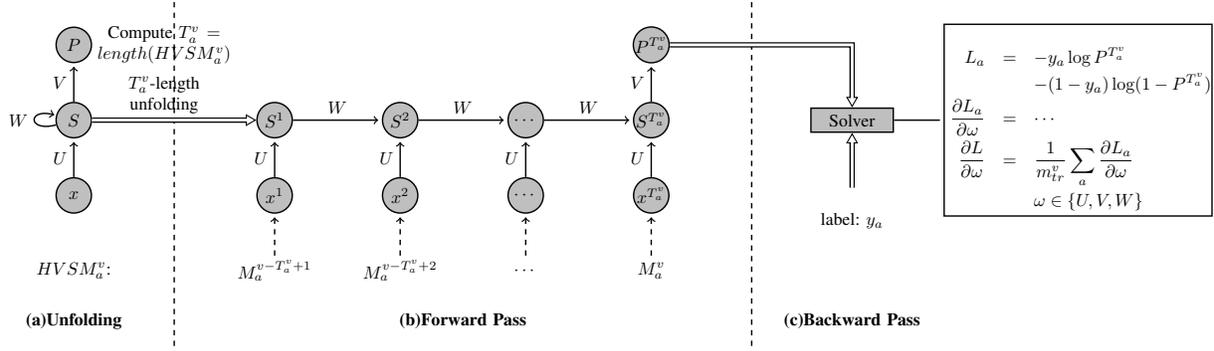
\begin{figure*}
	
	\centering
	\resizebox{0.9\textwidth}{!}{
		\begin{tikzpicture}[shorten >=2pt, auto, node distance=\layersep,thick]
		\tikzstyle{every pin edge}=[<-,shorten <=1pt]
		\tikzstyle{neuron}=[circle,fill=black!25,minimum size=20pt,inner sep=0pt]
		\tikzstyle{hidden neuron}=[neuron, draw=black];
		\tikzstyle{output neuron}=[neuron, draw=black];
		\tikzstyle{input neuron}=[neuron, draw=black];
		\tikzstyle{input metric}=[neuron, fill=white];
		\tikzstyle{annot} = [text width=8em, text centered,minimum size=20pt,];
		\tikzstyle{vecArrow} = [thick, decoration={markings,mark=at position
			1 with {\arrow[semithick]{open triangle 60}}},
		double distance=1.4pt, shorten >= 5.5pt,
		preaction = {decorate},
		postaction = {draw,line width=1.4pt, white,shorten >= 4.5pt}]
		\newcommand{\first}{-5}
		\newcommand{\unfolding}{-1.5}
		\newcommand{\forward}{2.5}
		\newcommand{\ydis}{-12 cm}

		% \node[annot] (Model) at (\first,2.5) {Model level};
		% \node[annot] (HVSM) at (\first,-2.5){Data level\\$HVSM_i$};

		\node[input neuron] (I) at (\unfolding,1) {$x$};
		\node[hidden neuron] (H) at (\unfolding,2.5) {$S$};
		\path (H) edge [loop left,->,] node{$W$} (H);
		\node[output neuron] (O) at (\unfolding,4) {$P$};
		\path (I) edge[->] node{$U$} (H);
		\path (H) edge[->]  node{$V$} (O);
		\node[annot] (Ti) at (\unfolding,-0.5){ $HVSM_a^v$:};
		\node[annot] at (\unfolding,-1.5) {\textbf{(a)Unfolding}};
		\draw[dashed] (0.5,-2) -- (0.5,5);
		\draw[dashed] (12,-2) -- (12,5);

		% \path (Ti) edge[-,shorten >=0pt,double distance=1.5 pt] (\unfolding + 2,-2.5);
		% \path (\unfolding + 2,-2.5)  edge[double distance=1.5 pt] (\unfolding + 2,2.5);

		% Draw the input layer nodes
		\foreach \name / \y in {1,2}
		\node[input neuron] (I-\name) at (\forward * \name ,1) {$x^{\name}$};
		\node[input neuron] (I-3) at (\forward * 3 ,1) {$\cdots$};
		\node[input neuron] (I-4) at (\forward * 4 ,1) {$x^{T_a^v}$};
		
		% Draw the hidden layer nodes
		
		\node[input metric] (M-1) at (\forward * 1 ,-0.5) {$M_a^{v-T_a^v+1}$};
		\node[input metric] (M-2) at (\forward * 2,-0.5) {$M_a^{v-T_a^v+2}$};
		\node[input metric] (M-3) at (\forward * 3 ,-0.5) {$\cdots$};
		\node[input metric] (M-4) at (\forward * 4,-0.5) {$M_a^{v}$};
		
		\foreach \name in {1,2}
		\node[hidden neuron] (H-\name) at (\forward * \name ,2.5 cm) {$S^{\name}$};
		\node[hidden neuron] (H-3) at (\forward * 3,2.5) {$\cdots$};
		\node[hidden neuron] (H-4) at (\forward * 4,2.5) {$S^{T_a^v}$};
		
		\path (H) edge[vecArrow] (H-1);
		
		\node[annot] at (0.3,4) {Compute $T_a^v=length(HVSM_a^v)$};
		\node[annot] at (0.3,3) {$T_a^v$-length unfolding };

		\foreach \source in {1,2,3,4}
		\path (I-\source) edge[->] node{$U$} (H-\source);
		
		\foreach \source in {1,2,3,4}
		\path  (\forward * \source,-0.2) edge[->,dashed]  (I-\source);
		\node[output neuron] (Output) at (10,4) {$P^{T_a^v}$};
		
		\path (H-4) edge[->] node{$V$} (Output);
		
		\foreach \pre / \next in {1/2,2/3,3/4}
		\path (H-\pre) edge[->] node{$W$} (H-\next);
		
		%\foreach \pre / \next in {1/2,2/3,3/4}
		%\path (M-\pre) edge[shorten >=8pt, shorten <=8pt,->] (M-\next);
		
		\node[annot,text width=10em] at (6.25,-1.5) {\textbf{(b)Forward Pass}};
		% \node[annot,text width=10em] at (6.25,-5) {\large{\textbf{Training }}};
		
		\node[input metric] (label) at (14,0.5) {label: $y_a$};
		\node[annot,text width=5.1cm,draw=black](computer) at (18.5,2.5) {
			\begin{eqnarray*}
			%Define\, &&\\
			L_a &=& -y_a\log P^{T_a^v}\\
			&&-(1-y_a)\log(1-P^{T_a^v})\\
			%Comput\, &&\\
			\frac{\partial L_a}{\partial \omega} &=&\cdots\\
			%Average\, &&\\
			\frac{\partial L}{\partial \omega} &=&\frac{1}{m_{tr}^v}\sum_a\frac{\partial L_a}{\partial \omega} \\
			&& \omega\in\{U,V,W\}
			\end{eqnarray*}
		};
		\node[rectangle,draw=black,text width=4em,text centered,fill=black!25](comparision) at (14,2.5) {Solver};
		\node[annot,text width=10em] at (14,-1.5) {\textbf{(c)Backward Pass}};
		\path (label) edge[vecArrow] (comparision);
		\draw[vecArrow] (Output)-- ++(4,0) --  (comparision);
		\path (comparision) edge (computer);
		\end{tikzpicture}
	}
	\caption{Training process of RNN on a HVSM}
	\label{ourrnn}
\end{figure*}

\subsection{Applying RNN to HVSM}
\label{sub:apply_rnn_to_hvsm}

In this section, we will introduce how to apply a proper classification technique, which is RNN in this paper, to the data of HVSMs to improve the performance of defect prediction.

In our study, for file $a$ in a given version $v$ of a project, the input of typical classifiers will be the metric set of $a$, i.e. $M_a^v$. When it comes to RNN, the input should be the HVSM of $a$, i.e. $HVSM_a^v$. It should be noticed that RNN can handle sequential data (see Section \ref{section.RNN}), so RNN will still work if the length of each file's HVSM differs in training set or test set.

%In our study, both the training set and test set of RNN should be in the form of a set of HVSM. For the project in Figure \ref{figure.HVSM}, we want to do the cross validation defect prediction on the file in version $v$, so we need to collect $HVSM^v$. For a file $a$ in version $v$, we search backward from version $v$ to find file $a$ in previous versions and then extract $HVSM_a^v$. After extracting the set $HVSM^{v}$, we split it into training set and test set and we apply RNN to HVSM.  As we have mentioned in Section \ref{section.RNN}, RNN can handle sequential data, so in this work, we just make the HVSM of a file be the input of RNN.

Figure \ref{ourrnn} illustrates the training process. First, we obtain a training sample $HVSM_a^v$ from the training set and then unfold RNN along the input sequence $HVSM_a^v$ according to its length $T_a^v$. Then, a forward pass is done: we let each metric set $M_a^{v-T_a^v+t}$ of $HVSM_a^v$ be the input of RNN at each time step, i.e.
\begin{equation}
x^t=M_a^{v-T_a^v+t}
\end{equation}
where $t \in \{1, 2, ..., T_a^v\}$ and then compute hidden state $S^1, S^2,$ $..., S^{T_a^v}$ successively according to equation \ref{hiddenstate} as well as obtain the probability that the file $a$ in version $v$ has bugs $P^{T_a^v}$ at last. Finally, in backward pass, we obtain the loss $L_a$ and compute its gradient $\frac{\partial L_a}{\partial \omega}$ on $HVSM_a^v$ with back-propagation technique, where $\omega$ represents the weight parameters $U_{ij}, V_{ij}, W_{ij}, b_k$ and $c$ in equation \ref{gd}. After obtaining gradient on all the training samples, we compute the gradient needed in equation \ref{gd} by
\begin{equation} 
\frac{\partial L}{\partial\omega}=\frac{1}{m^v_{tr}}\sum_{a}\frac{\partial L_a}{\partial\omega}
\end{equation}
where $m^v_{tr}$ is the number of training samples. Then the new weights $U, V, W, b$ and $c$ can be obtained by the iterations in equation \ref{gd}.

Once we complete the training process, predictions on the test set could be made by a similar process: for any instance $HVSM_b^{v}$ in the test set, first do a $T_b^{v}-$length unfolding, and then compute the probability $P^{T_b^{v}}$ that file $b$ in version $v$ has bugs via forward pass.

The most important thing to note is that in the RNN model, the same weights are reused at every time step (parameters sharing), which makes our RNN able to handle HVSMs with variable length. For example, during training process, for file $\heartsuit$ in version $v$, $T_{\heartsuit}^v=length(HVSM_{\heartsuit}^v)=3$, then we get a $3-$length unfolding RNN to process it. During the prediction process, for file $\diamondsuit$ in version $v$, $T_{\diamondsuit}^{v}=length(HVSM_{\diamondsuit}^{v})=2$, then we get a $2-$length unfolding RNN to process it.

%ourrnn
\section{Experimental Setup}
\label{section.experiment}
We conduct several experiments to study the performance of RNN using HVSMs and compare it with existing classification techniques.
\subsection{Collected Datasets}
We use data from 9 projects in the PROMISE data repository to make it available to replicate and verify our experiment. Each project has several versions and code metrics with clear defect information that label the buggy files. As shown in Table \ref{table.dataset}, the length of each project's version sequence varies from 3 versions to 5 versions, which makes it available to extract HVSM with $len>1$ when predicting WPDP. The data of these projects was investigated and shared by Jureczko and Madeyski \cite{Promise.data0} and also used by others' work \cite{technology.loc,Promise.data1,SementicFeatures}. 

%The static metrics here refer to the 20 metrics \cite{Promise.data0} investigated by Jureczko. 

%Table \ref{table.dataset} shows a clear view of the datasets we use. For each project, we choose the last pair of versions in the version sequence to extract their HVSMs as our experiment target. For example, the project \textbf{ant} has a long version sequence from version 1.3 to version 1.7. According to its last two version 1.6 and 1.7, we construct two HVSMs $HVSM^{1.6}$ and $HVSM^{1.7}$,  which will be used in the following experiment as training set and test set for RNN respectively. 

%In Table 1, we also summarize the distribution of three types  (developing, newborn and death) of modules discussed in section 3.1.  We can see that the module death rates are below 50\% for all the data sets and 80\% of them  lower than 20\%. The ratio of developing, newborn and dead modules in the studied versions of all the projects are 62.9\%, 25.2\% and 11.9\% respectively. Since the HVSM includes the code metrics of developing modules and newborn modules, we can say that the HVSMs here have covered most (88.1\%) of the modules in the projects' lifetime.

% Table generated by Excel2LaTeX from sheet 'Sheet2'
\begin{table}[htbp]
	\centering
	\caption{Information of collected projects}
	\label{table.dataset}%
	\begin{tabular}{l|l|c|c|c}
		\toprule
		Project & Collected Versions  & \tabincell{c}{Avg.\\\#Files} & \tabincell{c}{Avg.\\KLOC} & \tabincell{c}{Avg.\\\%Bugs}  \\
		\midrule
		ant   & 1.3, 1.4, 1.5, 1.6, 1.7 & 338   & 100   & 19.6\% \\
		camel & 1.0, 1.2, 1.4, 1.6 & 696   & 78    & 18.9\% \\
		jedit & 3.2, 4.0, 4.1, 4.2, 4.3 & 350   & 160   & 19.6\% \\
		log4j & 1.0, 1.1, 1.2 & 150   & 27    & 50.4\% \\
		lucene & 2.0, 2.2, 2.4 & 261   & 72    & 54.9\% \\
		poi   & 1.5, 2.0, 2.5, 3.0 & 345   & 99    & 49.8\% \\
		velocity & 1.4, 1.5, 1.6 & 213   & 54    & 58.5\% \\
		xalan & 2.4, 2.5, 2.6 & 804   & 314   & 36.6\% \\
		xerces & init, 1.2, 1.3, 1.4 & 411   & 140   & 38.3\% \\
		\bottomrule
	\end{tabular}%
\end{table}%

\subsection{Baseline Classifiers}
In this work, 7 typical classifiers are used as baselines to compare the performance with our RNN techniques. Most of the typical classifiers are studied by previous work \cite{Elish2008Predicting,Sun2012Using,Wang2010Naive,Rahman2013,Panichella:2016:STA:2908812.2908938} in defect prediction. The 7 classifiers are Naive Bayes (NB), Logistic Regression (LR), k-Nearest Neighbor (KNN), Random Forest (RF), C5.0 decision tree (C5.0), standard Neural Network (NN) and C4.5-like decision trees (J48). 
%These classifiers cover a wide range of classification technology including decision-tree, bagging, boosting, neural network, etc. 

\subsection{Model Construction}
\label{section.modelConstruction}

%Table \ref{table.dataset} shows a clear view of the datasets we use. For each project, we choose the last pair of versions in the version sequence to extract their HVSMs as our experiment target. For example, the project \textbf{ant} has a long version sequence from version 1.3 to version 1.7. According to its last two version 1.6 and 1.7, we construct two HVSMs $HVSM^{1.6}$ and $HVSM^{1.7}$,  which will be used in the following experiment as training set and test set for RNN respectively. 

%In Table 1, we also summarize the distribution of three types  (developing, newborn and death) of modules discussed in section 3.1.  We can see that the module death rates are below 50\% for all the data sets and 80\% of them  lower than 20\%. The ratio of developing, newborn and dead modules in the studied versions of all the projects are 62.9\%, 25.2\% and 11.9\% respectively. Since the HVSM includes the code metrics of developing modules and newborn modules, we can say that the HVSMs here have covered most (88.1\%) of the modules in the projects' lifetime.

For each project, we have 3 to 5 collected versions, which construct the version sequence of this project.  In order to observe the performance of our HVSM, the studied versions are those with enough number of previous versions, which provide HVSM with enough length.  

\begin{table}[htbp]
	\centering
	\caption{An example of metrics and versions included for typical classifiers and RNN}
	\label{table.modelConstruction}%
	\resizebox{1\columnwidth}{!}{
	\begin{threeparttable}
		\begin{tabular}{c|l|l|l}
			\toprule
			&       & RNN   & \multicolumn{1}{l}{Typical Classifiers} \\
			\midrule
			\multirow{2}[0]{*}{Training Set} & Metric set & $HVSM^{1.6}$ & $M^{1.6}$ \\
			& Versions \tnote{1} & ant 1.3, 1.4, 1.5, 1.6 &ant 1.6 \\
			\midrule
			\multirow{2}[0]{*}{Test Set} & Metric set & $HVSM^{1.7}$ &$M^{1.7}$ \\
			& Versions &ant 1.3, 1.4, 1.5, 1.6, 1.7 &ant 1.7 \\
			\bottomrule
		\end{tabular}%
		\begin{tablenotes}
			\footnotesize
			\item[1] refers to the versions included in the corresponding metric set
		\end{tablenotes}
	\end{threeparttable}
}
\end{table}%

\begin{table}[htbp]
	\centering
	\caption{An overview of HVSMs extracted as training and test set in each project }
	\label{table.selectedDataset}%
	\resizebox{1\columnwidth}{!}{
		\begin{threeparttable}
			\begin{tabular}{l|c|l|c|c|c|c}
				\toprule
				Project (Tr-$>$T)\tnote{1} & HVSM  & \tabincell{l}{Version\\Sequence} & \tabincell{c}{Avg.\\ $len$ \tnote{2}} & \#Files & \#DF\tnote{3}  & \%DF \\
				\midrule
				\multirow{2}[0]{*}{ant 1.6-$>$1.7} & $HVSM^{1.6}$ & 1.3,1.4,1.5,1.6 & 2.6   & 351   & 293   & 83.5\% \\
				& $HVSM^{1.7}$ & 1.3,1.4,1.5,1.6,1.7 & 2.3   & 745   & 355   & 47.7\% \\
				\midrule
				\multirow{2}[0]{*}{camel 1.4-$>$1.6} & $HVSM^{1.4}$ & 1.0,1.2,1.4 & 2.0   & 872   & 577   & 66.2\% \\
				& $HVSM^{1.6}$ & 1.0,1.2,1.4,1.6 & 2.7   & 965   & 857   & 88.8\% \\
				\midrule
				
				\multirow{2}[0]{*}{jedit 4.2-$>$4.3} & $HVSM^{4.2}$ & 3.2,4.0,4.1,4.2 & 3.2   & 367   & 291   & 79.3\% \\
				& $HVSM^{4.3}$ & 3.2,4.0,4.1,4.2,4.3 & 2.4   & 492   & 225   & 45.7\% \\
				\midrule
				\multirow{2}[0]{*}{log4j 1.1-$>$1.2} & $HVSM^{1.1}$ & 1.0,1.1 & 1.9   & 109   & 98    & 89.9\% \\
				& $HVSM^{1.2}$ & 1.0,1.1,1.2 & 2.0   & 205   & 117   & 57.1\% \\
				\midrule
				\multirow{2}[0]{*}{lucene 2.2-$>$2.4} & $HVSM^{2.2}$ & 2.0,2.2 & 1.8   & 247   & 192   & 77.7\% \\
				& $HVSM^{2.4}$ & 2.0,2.2,2.4 & 2.2   & 340   & 235   & 69.1\% \\
				\midrule
				\multirow{2}[0]{*}{poi 2.5-$>$3.0} &$HVSM^{2.5}$ & 1.5,2.0,2.5 & 2.4   & 385   & 314   & 81.6\% \\
				& $HVSM^{3.0}$ & 1.5,2.0,2.5,3.0 & 3.1   & 442   & 382   & 86.4\% \\
				\midrule
				\multirow{2}[0]{*}{velocity 1.5-$>$1.6} & $HVSM^{1.5}$ & 1.4,1.5 & 1.7   & 214   & 155   & 72.4\% \\
				& $HVSM^{1.6}$ & 1.4,1.5,1.6 & 2.6   & 229   & 210   & 91.7\% \\
				\midrule
				\multirow{2}[0]{*}{xalan 2.5-$>$2.6} & $HVSM^{2.5}$ & 2.4,2.5 & 1.9   & 803   & 689   & 85.8\% \\
				& $HVSM^{2.6}$ & 2.4,2.5,2.6 & 2.6   & 885   & 766   & 86.6\% \\
				\midrule
				\multirow{2}[0]{*}{xerces 1.3-$>$1.4} &$HVSM^{1.3}$ & init,1.2,1.3 & 2.2   & 453   & 433   & 95.6\% \\
				& $HVSM^{1.4}$ & init,1.2,1.3,1.4 & 2.2   & 588   & 328   & 55.8\% \\
				\bottomrule
			\end{tabular}%
			\begin{tablenotes}
				\footnotesize
				\item[1] Tr denotes the training set and T denotes the test set.
				\item[2] refers to $length(HVSM)$ (see Section \ref{section.extracting}).
				\item[3] refers to Developing Files. For a given version $v$, developing files were created before version $v$, and still exist in $v$ (see Section \ref{section.HVS}).
				
			\end{tablenotes}
		\end{threeparttable}
	}
	
\end{table}%

In WPDP, files in an older version $a$ are used as training set to predict defects in a newer version $b$, and the two versions are in the same project. For the metrics used in training set and test set, typical classifiers should perform defect prediction using $M^v$ in version $v$ while RNN uses $HVSM^v$. For example, considering the situation that researchers want to predict defects in \textbf{ant 1.7} with extracted metrics and labeled bugs in \textbf{ant 1.6} and former versions. Table \ref{table.modelConstruction} lists the metrics used by typical classifiers and our approach in this situation. Typically, classifiers use metrics in the training set \textbf{ant 1.6} without historical information in former versions, while our approach RNN use $HVSM^{1.6}$ which contains the sequential information from \textbf{ant 1.3} to \textbf{ant 1.6} as training data. In order to include more historical information, this study selects the \textbf{last two versions} in each project's version sequence as the training and test set in WPDP, and make the length of HVSM long enough. Table \ref{table.selectedDataset} shows the extracted HVSMs of the two selected versions as training and test set in WPDP in each project. The table also shows the number and ratio of developing files in each selected version. The high average percentage of developing files insures that HVSM will cover most of the files in a project's version history.

In order to train our defect prediction models, we use the implementations of the 7 typical classifiers provided by R packages. RNN is manually implemented using MATLAB since there is no suitable RNN packages with the data in the forms of HVSM as the input. We use the MATLAB code \textbf{fming.m} written by Carl Edward Rasmussen\footnote{http://learning.eng.cam.ac.uk/carl/} to realize the iteration process. To be noticed that some classifiers, like RF and NN based classifiers (RNN and NN) may have randomness in prediction, this paper repeat the classification process 10 times for each of these classifiers and report the mean value.  

This paper also applies the automated hyper-parameter optimization on the classification techniques introduced by Tantithamthavorn et al. \cite{Rparameter} using \texttt{caret R} \cite{caretR} package. 
%In this work, \texttt{caret} evaluates each candidate hyper-parameter settings by way of cross-validation in each training set. It then suggests the setting with the highest performance of AUC as the optimized one. For RNN, there are 3 hyper-parameters, i.e. the number of nodes in hidden layer $n_h$, regularization parameter $\lambda$ and the number of iterations $l_{total}$ which could affect the performance of RNN. 
%Since RNN is not included in the caret's model list, we imply the hyper-parameter optimization manually in the same way \textbf{caret} does on other techniques. 

\subsection{Metrics}
In our study we use code metrics and process metrics for our classification techniques and HVSM that RNN uses.
\subsubsection{Code Metrics}
Code metrics used for classification techniques in this study are extracted and investigated by Jureczko and Madeyski \cite{Promise.data0} in previous work. According to the paper, there are in total 20 code metrics including the common used LOC (lines of code) in addition to the other 19 metrics suggested by Chidamber and Kemerer \cite{metrics.CK}, Henderson-Sellers \cite{metrics.henderson}, Bansiy and Davis \cite{Bansiya2002}, Tang et al. \cite{Tang1999}, Martin \cite{Marting1994}, and McCabe \cite{metrics.McCabe}. 

\subsubsection{Process Metrics}
Different from code metrics which are static metrics within each release of projects, process metrics measure the change information of files during a period of time. In this study, 4 change metrics studied by Nagappan and Ball \cite{metric.codechurn} are extracted as process metrics. \textbf{ADD} and \textbf{DEL} measure the lines of code added or deleted in a specific file from last release to current release. \textbf{CADD} and \textbf{CDEL} are cumulative lines of code added or deleted during the whole version sequence until a specific release. These metrics are studied as effective predicting indicators by previous works \cite{Mockus2000,metric.codechurn,Moser2008A}. 

In total, there are 20 code metrics and 4 process metrics involved in this study. 

\subsection{Evaluation}

There are many evaluation measures in the field of defect prediction. In this study, we mainly focus on the performance of classification techniques in effort-aware scenarios. 
%It is suggested by Jiang et al. \cite{multiMetrics.summary} that the performance of defect prediction should be discussed under different evaluation metrics. In this work, we introduce 3 metrics to evaluate the performance. 

\subsubsection{Cost-Effectiveness}
\begin{figure}
	\centering
	\includegraphics[width=0.35\textwidth]{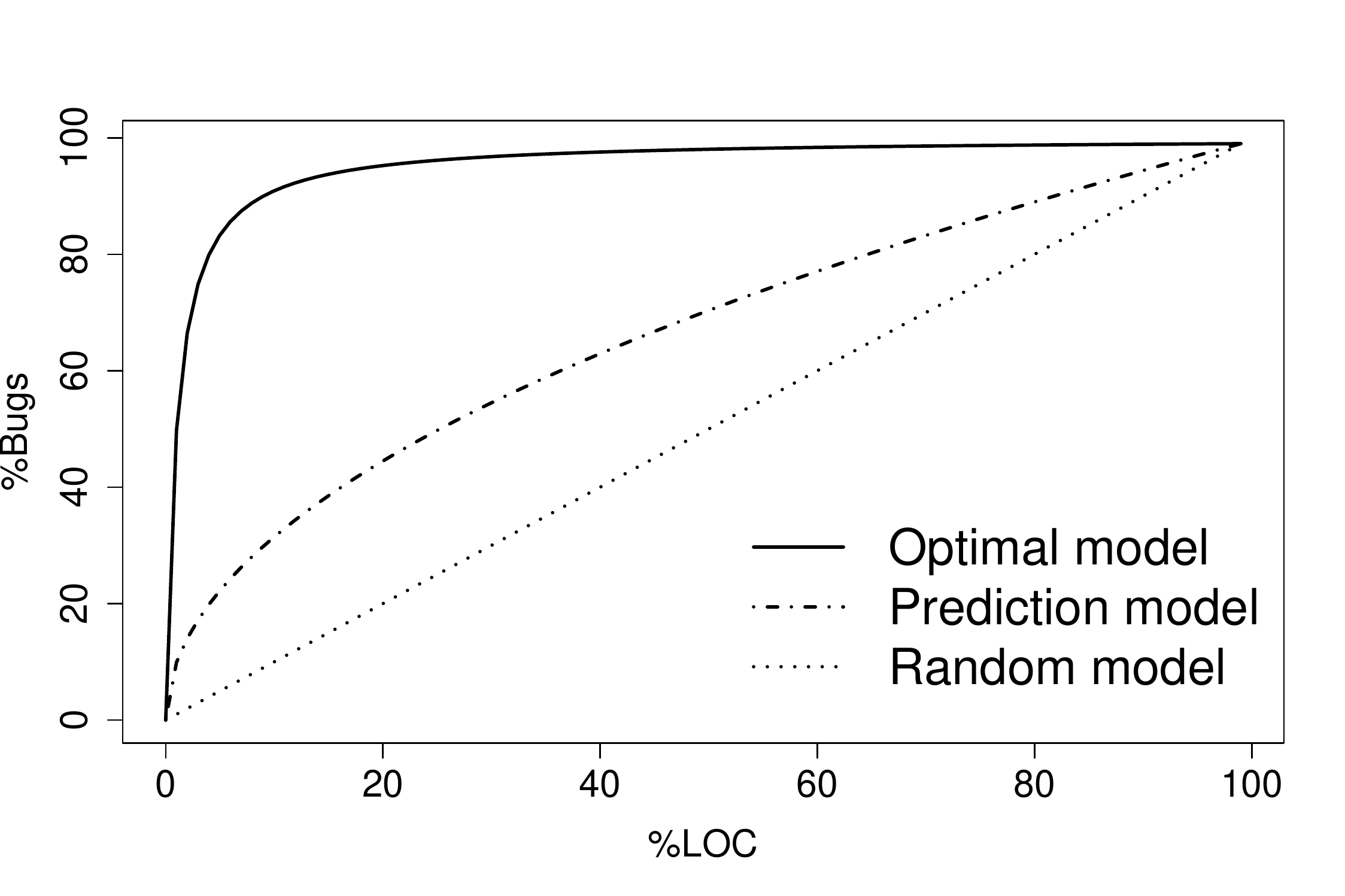}
	\caption{An example of CE curve}
	\label{ceplot}
\end{figure}
%It is widely used in prior works Both AUC and F1 are common metrics that are used to evaluate the performance of a classification technique. However, in effort-aware scenarios, we are most willing to rank the files in descending orders of bug density (buggy probability/size). In this situation, engineers can allocate less effort to more buggy files. 
When performing defect prediction, classifiers always rank the files by their probability of being defective. Sometimes practitioners do not have enough resources to inspect the whole project, they prefer to check those files with small size and high fault-proneness. In this situation, a cost-effective model would rank the files in descending order of their bug density. The effort-aware ranking effectiveness of classification techniques in defect prediction is always evaluated by cost-effectiveness (CE) curve, which is widely used in prior works \cite{Yang2016,Rahman2013,multiMetrics.MWY}. Figure \ref{ceplot} shows an example of the CE curve. In the figure, x-axis represents the cumulative percentage of LOC of the files, and y-axis is the cumulative percentage of bugs detected by the selected files. For a prediction model $A$, we sort the files in descending orders of $P(buggy)/LOC$, where $P(buggy)$ is the predicted probability of a file being defective. A CE curve of model $A$ plots proportion of defects truly detected against proportion of LOC coming from the ordered set of files. We use the following formula introduced by Arisholm et al. \cite{Arisholm2010A} to calculate CE :
\begin{equation*}
CE_\pi=\frac{Area_\pi(M)-Area_\pi(Random)}{Area_\pi(Optimal)-Area_\pi(Random)}
\end{equation*}
Where $Area_\pi(A)$ is the area under the curve of model $A$ ($M$, $Random$ or $Optimal$) for a given $\pi$ percentage of LOC. In random model, files are randomly selected to inspect, while in optimal model, files are ranked in descending order according to their actual bug densities. The larger $CE_\pi$ means a better ranking effectiveness. The cut-off $\pi$ varies from 0 to 1 indicating the percentage of LOC that we inspect. In this work, we report $CE_\pi$ at $\pi$ = 0.1, 0.2, 0.5 and 1.0.

\subsubsection{Scott-Knott Test}
Scott-Knott (SK) test \cite{sk.origin} (using the 95\% confidence level) is also applied in this paper to group classifiers into statistically distinct ranks. The SK test recursively ranks the evaluated classifiers based on hierarchical clustering analysis. It clusters the evaluated classifiers into two groups based on evaluation metrics, and recursively executes within each rank until no significant distinct groups can be created \cite{background.dp5}. The SK test has been used in prior works \cite{sk.apply1,sk.apply2,background.dp5,dp.Unsupervised} to compare the performance of different classifiers.

\subsubsection{Win/Tie/Loss}
%When comparing the performance of classification techniques on several projects, a common way is to calculate the average of a measure over all the projects. 
To further compare the performance of classifiers apart from average value or average rank, we also apply Win/Tie/Loss results which is also used for performance comparison between different techniques by prior works \cite{wintieloss2,background.wpdp,wintieloss1}. For each project, we repeat the model training of RNN and other techniques that have randomness 10 times and have 10 scores of the performance for each technique. For each technique that has no randomness (like logistic regression), the result is copied 10 times to make it comparable with RNN. After that, Wilcoxon signed-rank test \cite{Wilcoxon} together with Cliff's delta $\delta$ \cite{Romano2006Exploring} is conducted to compare the performance of RNN and other classification techniques. For a baseline technique A, if RNN outperforms A in a given project according to the Wilcoxon signed-rank test ($p<0.05$), and the magnitude of the difference between RNN and A is not negligible according to Cliff's delta $\delta$ ($\delta \geq 0.147$), we mark the RNN as a `Win'. In contrast, RNN is marked as a `Loss' if a technique B outperforms RNN in a project with statistical significance ($p<0.05$ and $\delta \geq 0.147$). Otherwise, the case is marked as a `Tie'. Finally, we count the Wins, Ties and Losses for RNN against each technique in each project. This Win/Tie/Loss evaluation shows that the number of projects in which RNN outperforms other techniques with statistical significance.

\section{Results}
\label{section.result}
This section provides our experimental results. We focus comparing our approach, RNN using HVSM, with other classifiers in within-project defect prediction (WPDP), and answer the following research question:
\subsection{RQ1: Does RNN with HVSM outperform other techniques in WPDP using code metrics? }
\label{section.rq1}
\begin{figure}
	\centering
	\subfigure[$CE_{0.1}$]{
		\label{figure.rq1_ce10} %% label for first subfigure
		\includegraphics[width=0.23\textwidth]{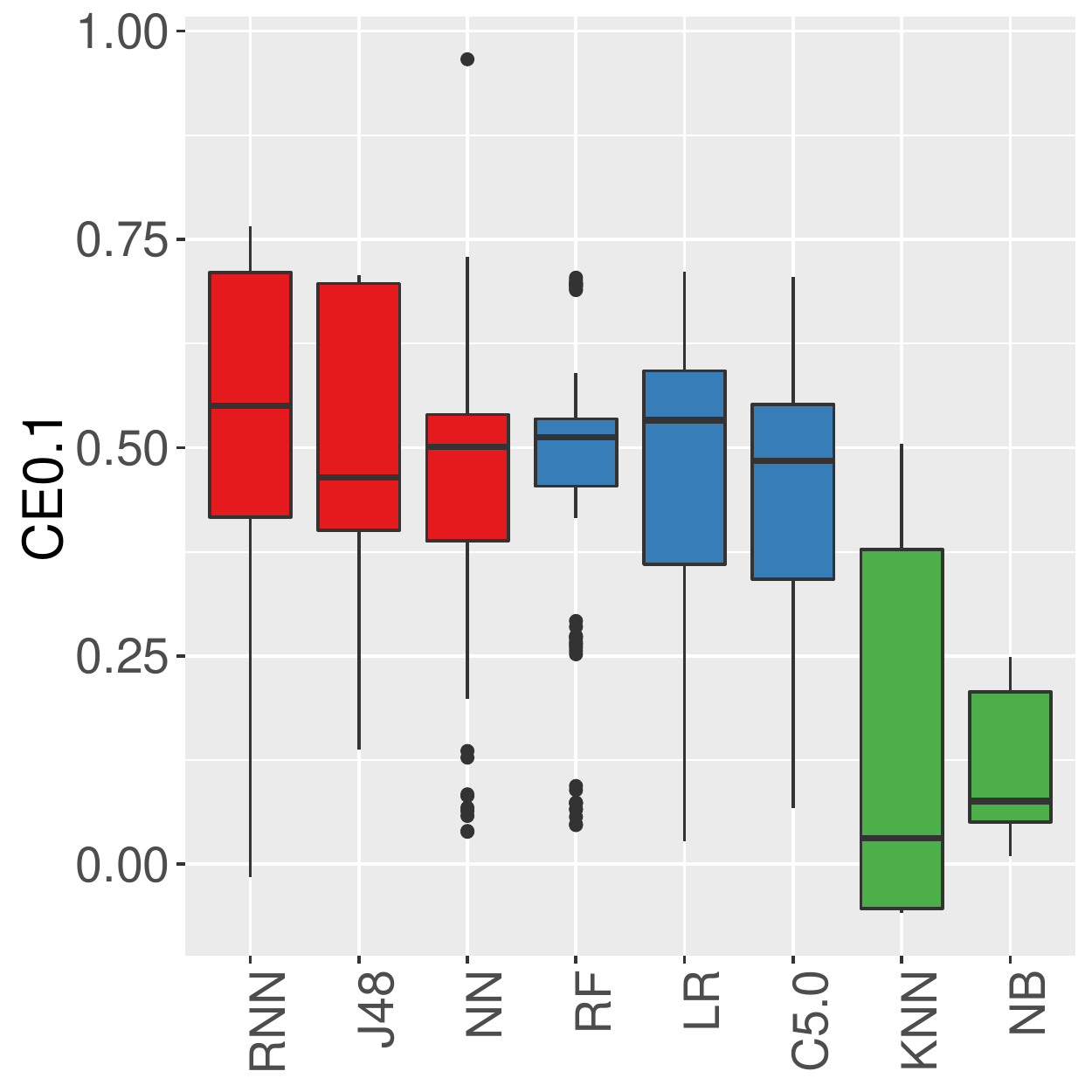}}
	\subfigure[$CE_{0.2}$]{
		\label{figure.rq1_ce20} %% label for first subfigure
		\includegraphics[width=0.23\textwidth]{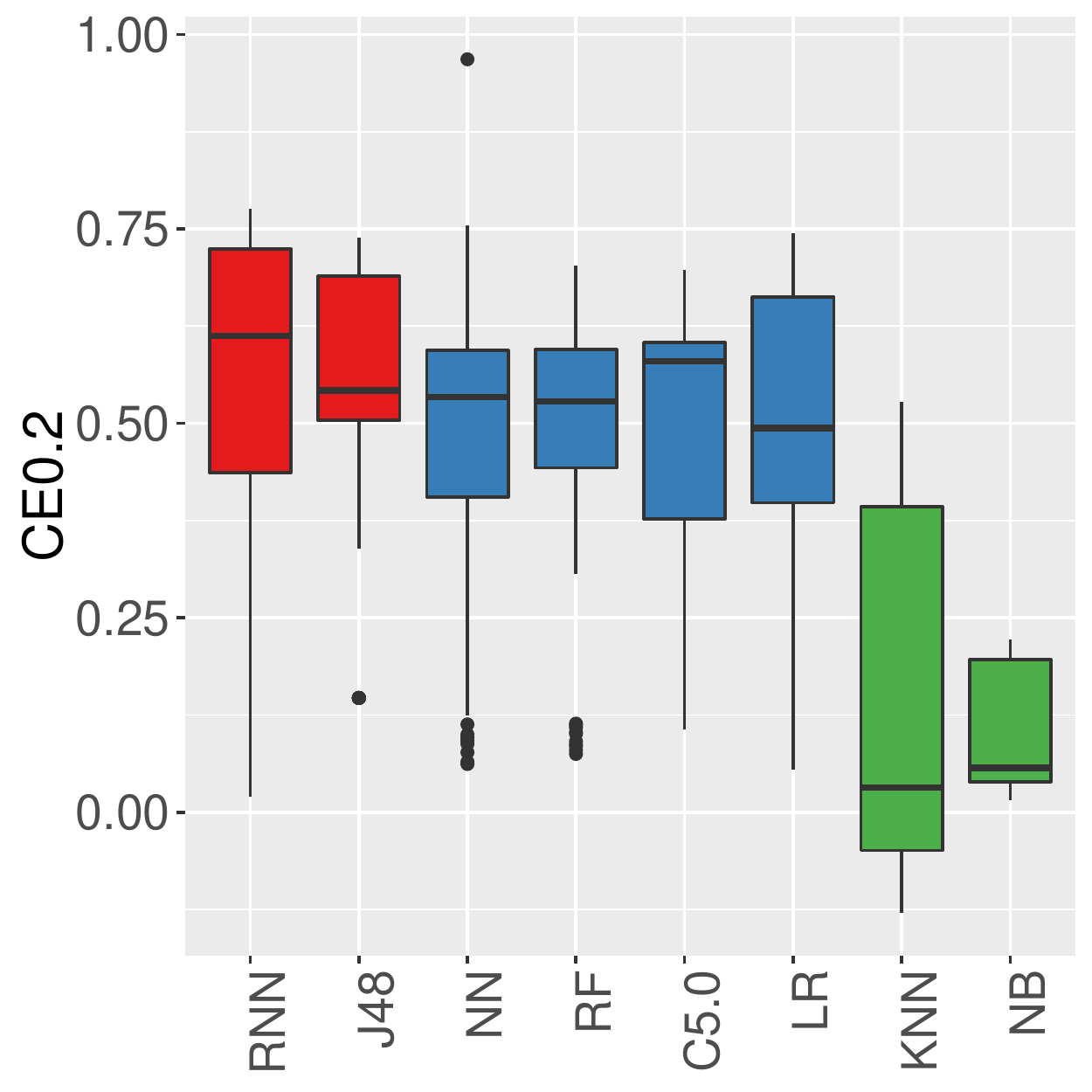}}
	\subfigure[$CE_{0.5}$]{
		\label{figure.rq1_ce50} %% label for first subfigure
		\includegraphics[width=0.23\textwidth]{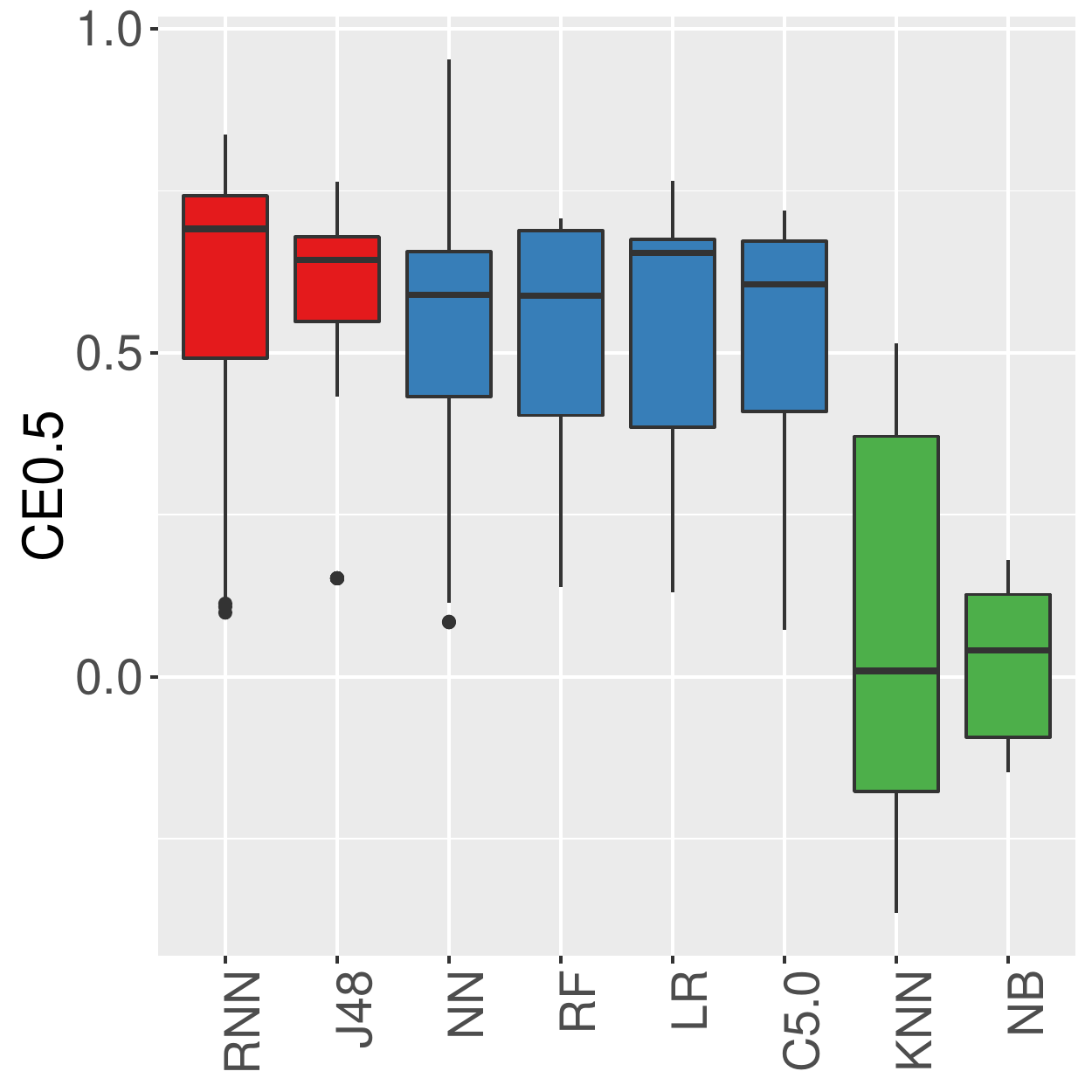}}
	\subfigure[$CE_{1.0}$]{
		\label{figure.rq1_ce100} %% label for first subfigure
		\includegraphics[width=0.23\textwidth]{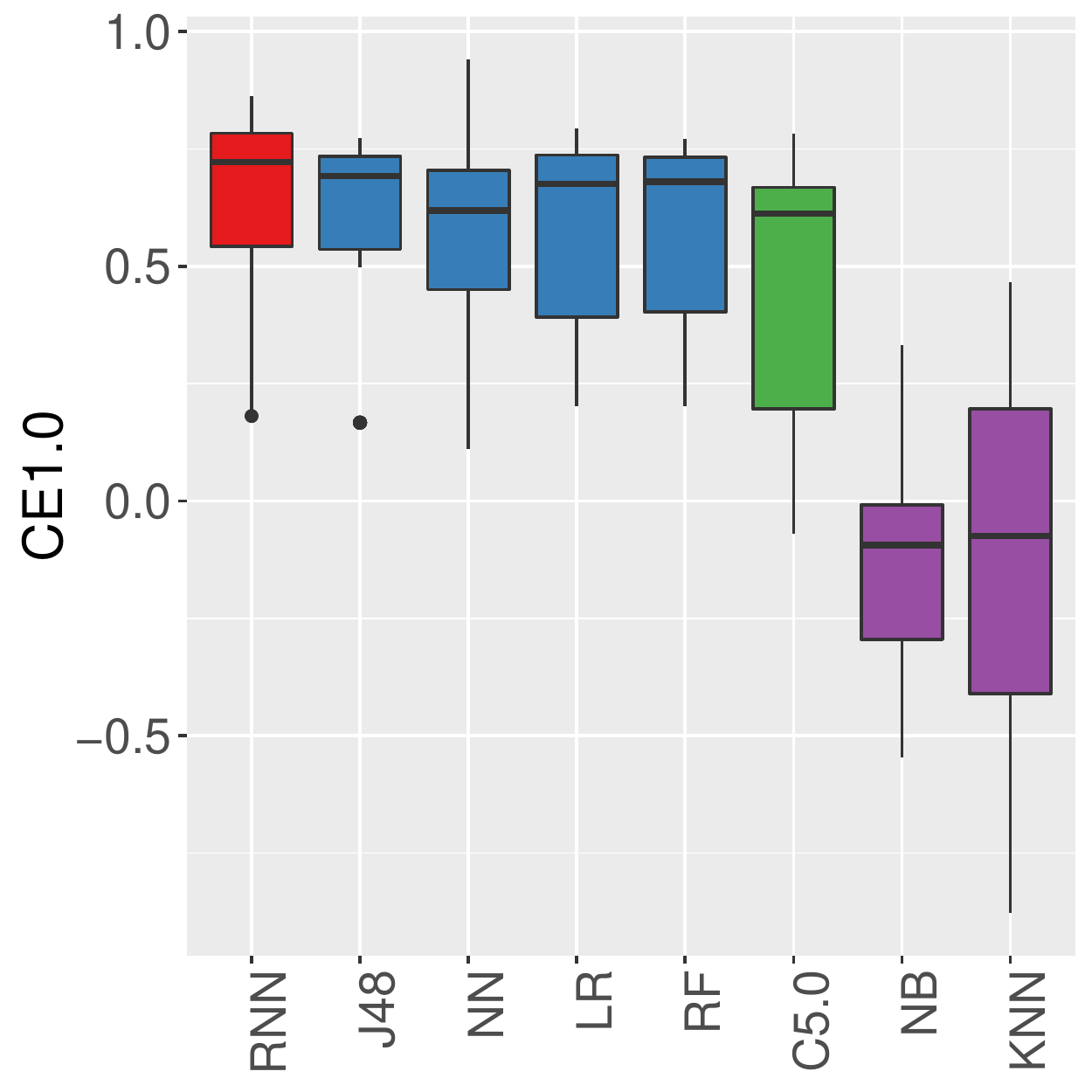}}	
	\caption{The boxplots of  $CE_{\pi}, \pi=0.1, 0.2, 0.5, 1.0$ values of RNN and 7 baseline classifiers using only code metrics. Different colors represents different Scott-Knott test ranks (from top down, the order is red, blue, green, purple).}
	\label{figure.rq1}
\end{figure}

\begin{figure}
	\centering
	\subfigure[$CE_{0.1}$]{
		\label{figure.rq2_ce10} %% label for first subfigure
		\includegraphics[width=0.23\textwidth]{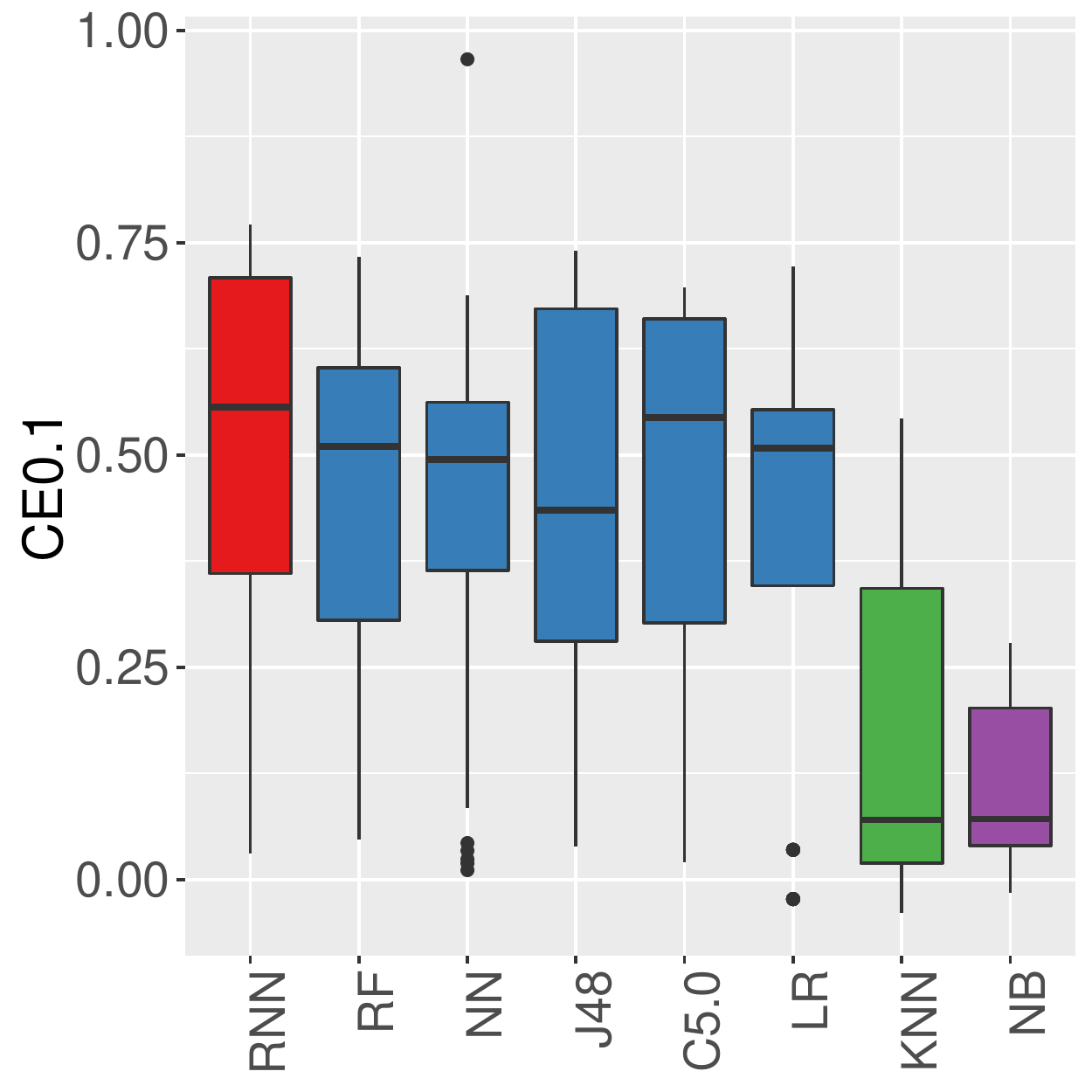}}
	\subfigure[$CE_{0.2}$]{
		\label{figure.rq2_ce20} %% label for first subfigure
		\includegraphics[width=0.23\textwidth]{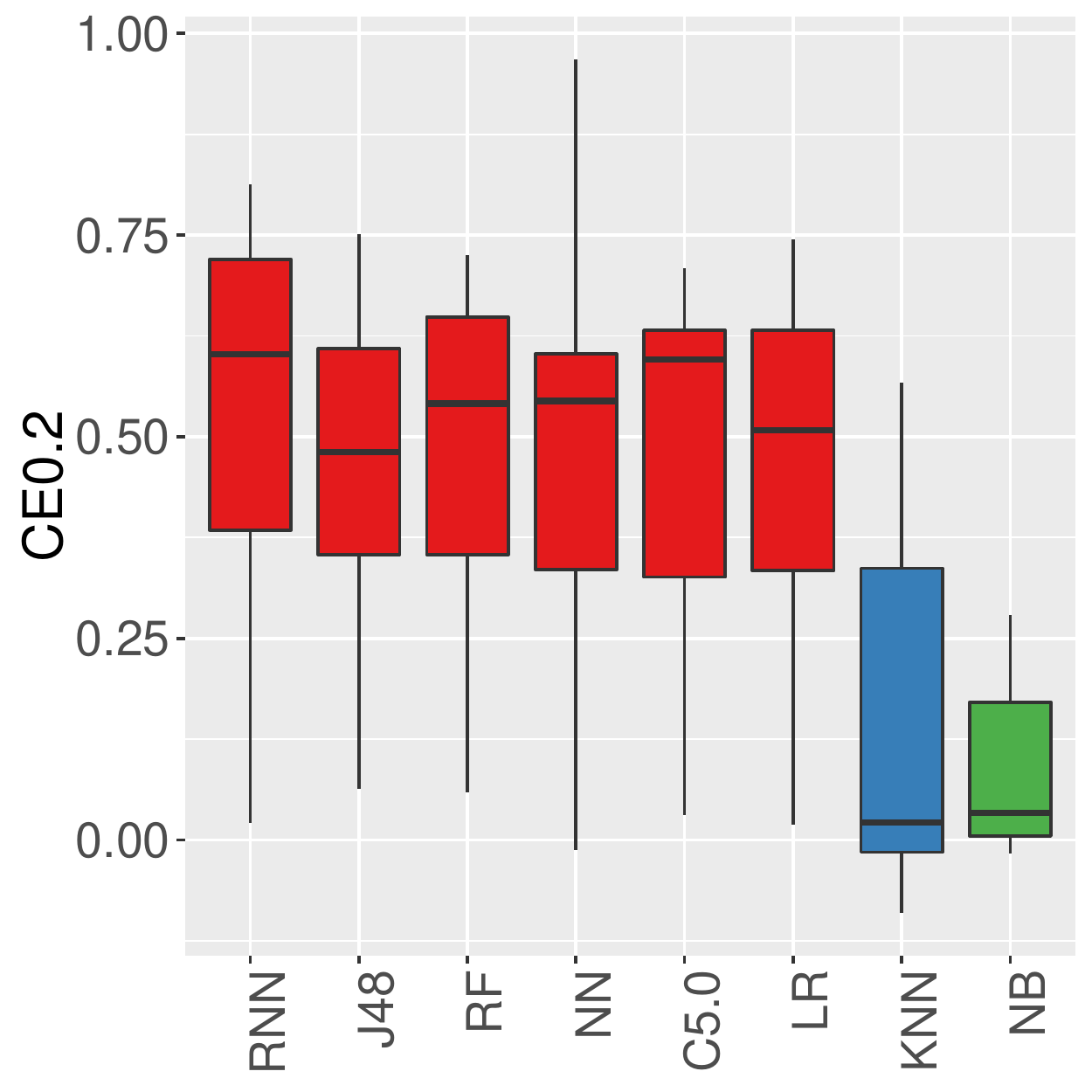}}
	\subfigure[$CE_{0.5}$]{
		\label{figure.rq2_ce50} %% label for first subfigure
		\includegraphics[width=0.23\textwidth]{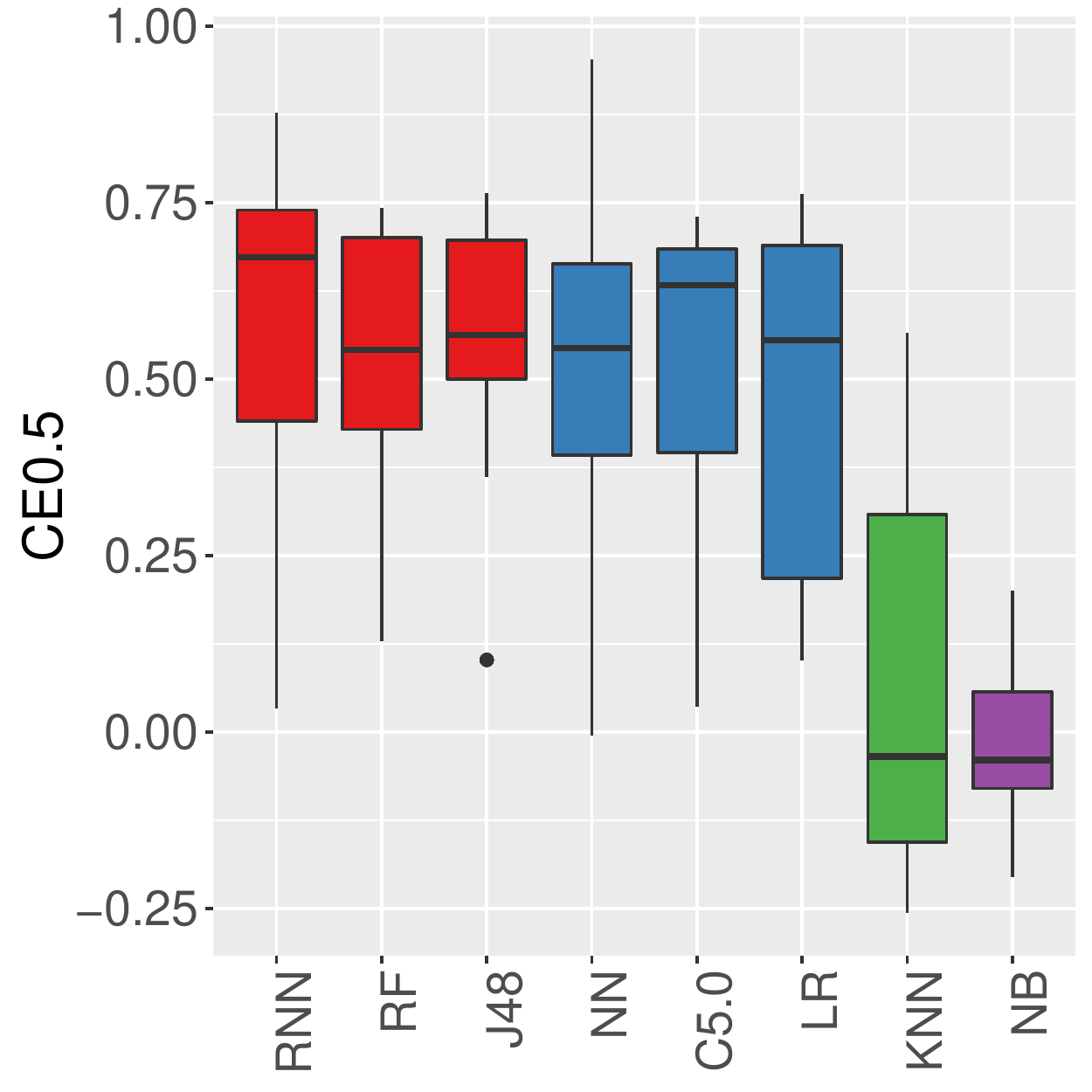}}
	\subfigure[$CE_{1.0}$]{
		\label{figure.rq2_ce100} %% label for first subfigure
		\includegraphics[width=0.23\textwidth]{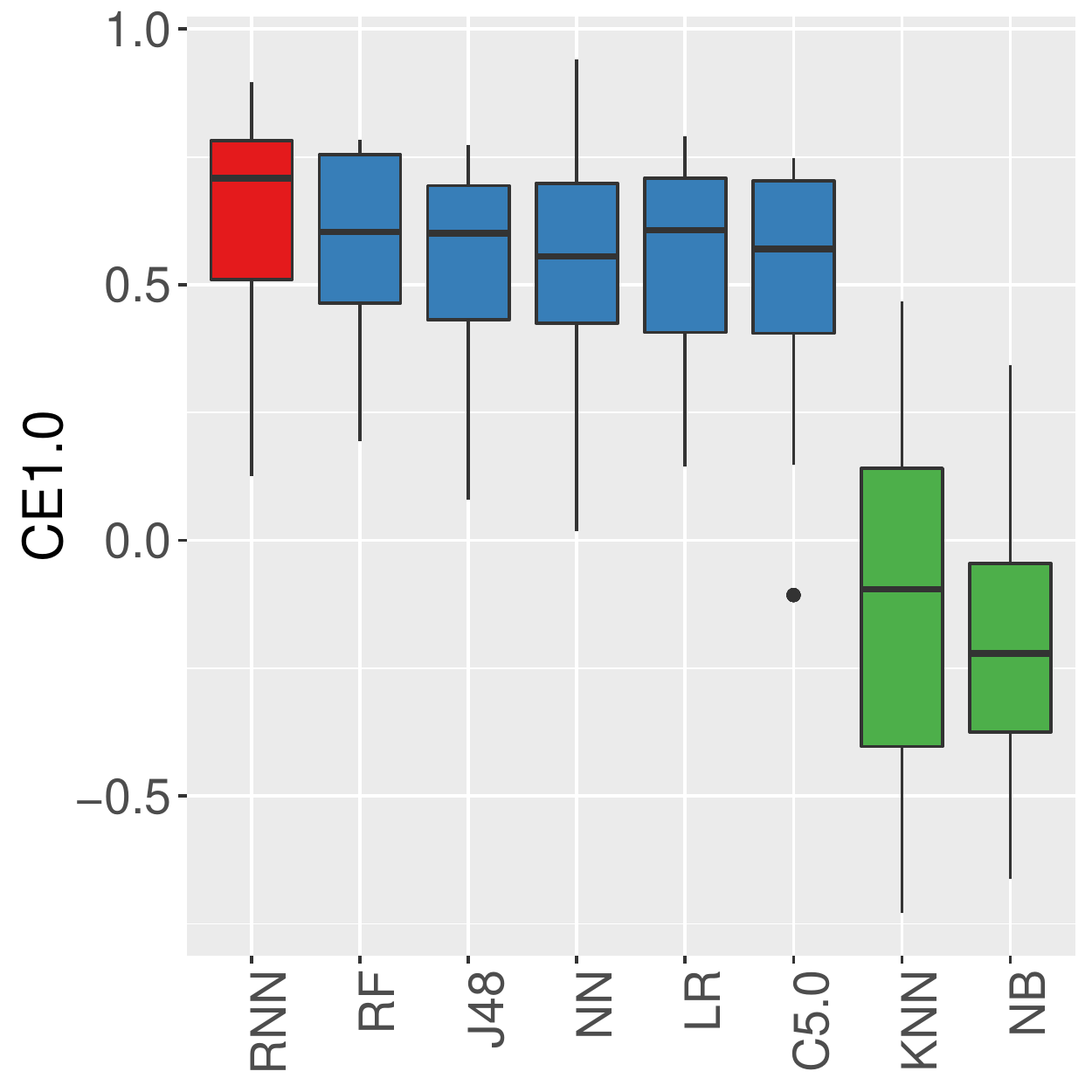}}	
	\caption{The boxplots of  $CE_{\pi}, \pi=0.1, 0.2, 0.5, 1.0$ values of RNN and 7 baseline classifiers using both code and process metrics. Different colors represents different Scott-Knott test ranks (from top down, the order is red, blue, green, purple).}
	\label{figure.rq2a}
\end{figure}

\begin{table*}
	\centering
	\caption{The $CE_{\pi}$ values of the top 4 classifiers using code metrics in WPDP. Bold font highlights the best performance. }
		\label{table.rq1}%
	\resizebox{1\textwidth}{!}{
	\begin{threeparttable}
		\begin{tabular}{l|cccc|cccc|cccc|cccc}
			\toprule
			& \multicolumn{4}{c|}{$CE_{0.1}$}      & \multicolumn{4}{c|}{$CE_{0.2}$}      & \multicolumn{4}{c|}{$CE_{0.5}$}      & \multicolumn{4}{c}{$CE_{1.0}$} \\
			\midrule
			Target (Tr-$>$T) & RNN   & J48   & NN    & RF    & RNN   & J48   & NN    & RF    & RNN   & J48   & NN    & RF    & RNN   & J48   & NN    & LR \\
			\midrule
			ant 1.6-$>$1.7 & 0.023 & \textbf{0.138} & 0.09  & 0.069 & 0.049 & \textbf{0.147} & 0.098 & 0.096 & 0.125 & 0.152 & 0.127 & \textbf{0.154} & \textbf{0.209} & 0.167 & 0.158 & 0.202 \\
			camel 1.4-$>$1.6 & 0.354 & 0.307 & \textbf{0.369} & 0.267 & \textbf{0.396} & 0.339 & 0.39  & 0.317 & \textbf{0.462} & 0.433 & 0.446 & 0.394 & \textbf{0.533} & 0.498 & 0.517 & 0.507 \\
			jedit 4.2-$>$4.3 & \textbf{0.438} & 0.401 & 0.391 & 0.46  & 0.455 & \textbf{0.508} & 0.372 & 0.479 & 0.496 & \textbf{0.643} & 0.378 & 0.584 & 0.537 & \textbf{0.734} & 0.43  & 0.392 \\
			log4j 1.1-$>$1.2 & \textbf{0.757} & 0.702 & 0.692 & 0.569 & \textbf{0.715} & 0.689 & 0.641 & 0.525 & \textbf{0.706} & 0.672 & 0.586 & 0.462 & \textbf{0.719} & 0.692 & 0.588 & 0.32 \\
			lucene 2.2-$>$2.4 & \textbf{0.757} & 0.707 & 0.705 & 0.696 & \textbf{0.767} & 0.711 & 0.73  & 0.699 & \textbf{0.781} & 0.721 & 0.752 & 0.706 & \textbf{0.806} & 0.731 & 0.784 & 0.794 \\
			poi 2.5-$>$3.0 & \textbf{0.6} & 0.463 & 0.505 & 0.511 & \textbf{0.67} & 0.579 & 0.575 & 0.62  & \textbf{0.737} & 0.568 & 0.637 & 0.687 & \textbf{0.781} & 0.536 & 0.681 & 0.779 \\
			velocity 1.5-$>$1.6 & \textbf{0.538} & 0.495 & 0.508 & 0.528 & \textbf{0.571} & 0.542 & 0.528 & 0.566 & 0.694 & 0.679 & 0.653 & \textbf{0.7} & \textbf{0.764} & 0.757 & 0.733 & 0.737 \\
			xalan 2.5-$>$2.6 & \textbf{0.548} & 0.464 & 0.514 & 0.533 & \textbf{0.612} & 0.504 & 0.582 & 0.593 & \textbf{0.671} & 0.548 & 0.648 & 0.664 & \textbf{0.708} & 0.537 & 0.686 & 0.692 \\
			xerces 1.3-$>$1.4 & \textbf{0.701} & 0.697 & 0.5   & 0.474 & \textbf{0.764} & 0.739 & 0.523 & 0.458 & \textbf{0.824} & 0.764 & 0.537 & 0.411 & \textbf{0.848} & 0.774 & 0.525 & 0.676 \\
			\midrule
			Avg.  & \textbf{0.524 } & 0.486  & 0.475  & 0.456  & \textbf{0.555 } & 0.529  & 0.493  & 0.484  & \textbf{0.611 } & 0.576  & 0.529  & 0.529  & \textbf{0.656 } & 0.603  & 0.567  & 0.567  \\
			AR    & 2.2   & 3.9   & 3.6   & 4.1   & 2.2   & 3.3   & 4.1   & 4.2   & 2.2   & 3.7   & 4.3   & 3.6   & 1.6   & 3.9   & 4.0   & 3.4 \\
			Win/Tie/Loss &   ---    & 7/1/1 & 7/1/1 & 7/0/2 &    ---   & 7/0/2 & 7/1/1 & 6/1/2 &    ---   & 7/0/2 & 7/2/0 & 6/0/3 &  ---     & 8/0/1 & 7/2/0 & 7/2/0 \\
			
			\bottomrule
		\end{tabular}%
		\begin{tablenotes}
			\footnotesize
			\item[1]  Tr denotes the training set version and T denotes the test set version.
		\end{tablenotes}
		
	\end{threeparttable}
}

\end{table*}%

\begin{table*}
	\centering
	\caption{The $CE_{\pi}$ values of the top 4 classifiers using both code and process metrics in WPDP. Bold font highlights the best performance.}
	\label{table.rq2a}%
	\resizebox{1\textwidth}{!}{
		\begin{tabular}{l|cccc|cccc|cccc|cccc}
			\toprule
			& \multicolumn{4}{c|}{$CE_{0.1}$}      & \multicolumn{4}{c|}{$CE_{0.2}$}      & \multicolumn{4}{c|}{$CE_{0.5}$}      & \multicolumn{4}{c}{$CE_{1.0}$} \\
			\midrule
			Target (Tr-$>$T) & RNN   & RF    & NN    & J48   & RNN   & J48   & RF    & NN    & RNN   & RF    & J48   & NN    & RNN   & RF    & J48   & NN \\
			\midrule
			ant 1.6-$>$1.7 & 0.064 & 0.056 & \textbf{0.068} & 0.039 & 0.054 & 0.064 & \textbf{0.072} & 0.064 & 0.089 & \textbf{0.14} & 0.102 & 0.082 & 0.165 & \textbf{0.221} & 0.08  & 0.115 \\
			camel 1.4-$>$1.6 & \textbf{0.348} & 0.302 & 0.334 & 0.281 & \textbf{0.377} & 0.293 & 0.334 & 0.34  & \textbf{0.442} & 0.418 & 0.361 & 0.396 & \textbf{0.51} & 0.48  & 0.371 & 0.456 \\
			jedit 4.2-$>$4.3 & \textbf{0.365} & 0.267 & 0.348 & 0.276 & \textbf{0.355} & 0.354 & 0.333 & 0.326 & 0.392 & 0.475 & \textbf{0.526} & 0.345 & 0.476 & 0.583 & \textbf{0.647} & 0.408 \\
			log4j 1.1-$>$1.2 & \textbf{0.737} & 0.612 & 0.619 & 0.672 & \textbf{0.721} & 0.609 & 0.558 & 0.604 & \textbf{0.707} & 0.537 & 0.589 & 0.55  & \textbf{0.709} & 0.516 & 0.601 & 0.549 \\
			lucene 2.2-$>$2.4 & 0.723 & 0.725 & 0.64  & \textbf{0.741} & 0.736 & \textbf{0.751} & 0.722 & 0.668 & \textbf{0.763} & 0.738 & 0.76  & 0.696 & \textbf{0.795} & 0.776 & 0.774 & 0.713 \\
			poi 2.5-$>$3.0 & \textbf{0.575} & 0.543 & 0.536 & 0.492 & \textbf{0.66} & 0.608 & 0.647 & 0.631 & 0.735 & \textbf{0.736} & 0.697 & 0.706 & 0.779 & \textbf{0.78} & 0.694 & 0.741 \\
			velocity 1.5-$>$1.6 & \textbf{0.545} & 0.508 & 0.506 & 0.435 & \textbf{0.559} & 0.481 & 0.542 & 0.526 & 0.638 & \textbf{0.679} & 0.562 & 0.602 & 0.72  & \textbf{0.753} & 0.569 & 0.685 \\
			xalan 2.5-$>$2.6 & 0.55  & \textbf{0.603} & 0.525 & 0.417 & 0.604 & 0.46  & \textbf{0.651} & 0.588 & 0.67  & \textbf{0.7} & 0.5   & 0.652 & 0.704 & \textbf{0.721} & 0.431 & 0.684 \\
			xerces 1.3-$>$1.4 & 0.69  & 0.495 & 0.485 & \textbf{0.697} & 0.737 & \textbf{0.739} & 0.498 & 0.507 & \textbf{0.811} & 0.439 & 0.764 & 0.53  & \textbf{0.833} & 0.23  & 0.774 & 0.52 \\
			\midrule
			Avg.  & \textbf{0.511 } & 0.457  & 0.451  & 0.450  & \textbf{0.534 } & 0.484  & 0.484  & 0.473  & \textbf{0.583 } & 0.540  & 0.540  & 0.507  & \textbf{0.632 } & 0.562  & 0.549  & 0.541  \\
			AR    & 1.7   & 3.5   & 4.1   & 4.0   & 2.1   & 3.6   & 3.1   & 4.1   & 2.2   & 2.6   & 3.6   & 4.6   & 1.7   & 2.4   & 4.3   & 4.3 \\
			Win/Tie/Loss &    ---   & 5/3/1 & 6/3/0 & 7/1/1 &---       & 5/3/1 & 6/2/1 & 6/3/0 &  ---     & 4/1/4 & 6/2/1 & 7/2/0 &  ---     & 4/1/4 & 8/0/1 & 7/2/0 \\
			
			\bottomrule
		\end{tabular}%
	}
	
\end{table*}%

Generally speaking, our approach outperforms other techniques in effort-aware scenarios evaluated by CE, and the result is supported by Scott-Knott test and Win/Tie/Loss results.  

Figure \ref{figure.rq1} shows an overview of our approach comparing with other classifiers. The boxplots show the distribution of $CE_\pi$ values of each classifier in the studied datasets. Different colors of the boxplot indicate different tiers that a classifier is ranked by SK test (using the 95\% confidence level). The SK result shows that our approach ranks the first (red boxplots) under all the evaluation metrics. 

Table \ref{table.rq1} shows the detailed comparison of $CE_\pi$ values of the top four techniques. Considering the average value, RNN has the best performance among the top four techniques under all the evaluation metrics. Highlighted by bold font, Our approach achieved the best performance in no less than 6 (out of 9) datasets evaluated by different $CE_\pi$. In addition, we also provide average rank (AR) \cite{background.dp1,dp.Unsupervised} of each technique over all the projects. AR can well reflect how a classifier outperforms others with little influence by extreme values in a few dataset which is also adopted by other works \cite{background.wpdp2,Lessmann.benchmark}. In the view of AR, our approach has 2.2, 2.2, 2.2 and 1.6 under $CE_{0.1}$, $CE_{0.2}$, $CE_{0.5}$ and $CE_{1.0}$ respectively, which are the best among the top 4 techniques. The superior AR shows a better applicability of our approach in different projects compared with other classifiers.

In order to further compare the classifiers, we also apply the Win/Tie/Loss indicator with the help of Wilcoxon signed-rank test and Cliff's delta $\delta$. The Win/Tie/Loss result shows whether RNN is significantly better or not when compared with other classifiers. Our approach achieves at least 6 `Win' and no more than 3 `Loss' against a specific classifier in all the cases, which means that RNN outperforms others in at least 6 (out of 9) datasets with statistical significance. This result supports the better performance of RNN compared with baseline classifiers.

\textbf{In summary, our approach outperforms baseline techniques using code metrics as training data in effort-aware scenarios evaluated by CE. The result is supported by SK test and  Win/Tie/Loss evaluation.}

\subsection{RQ2: Does our approach outperform other techniques in WPDP using both code and process metrics? }
\label{section.rq2}

In addition to code metrics, process metrics are also effective predictor in defect prediction. This paper also provides performance of RNN and typical classifiers using both code and process metrics as training data.

\subsubsection{RQ2a: How is the performance of RNN and typical classifiers using both code and process metrics?}
\label{section.rq2a}
In this section, the training data for typical classifiers includes 24 metrics (20 code metrics and 4 process metrics), and in HVSM used by RNN, metric set in each version also consists of these 24 metrics.

Similar with RQ1, Figure \ref{figure.rq2a} shows an overview of RNN comparing with other classifiers. Evaluated by $CE_{0.1}$, $CE_{0.5}$ and $CE_{1.0}$, our approach ranks the first and has significant distinction with most of the typical techniques. When it comes to $CE_{0.2}$, RNN still ranks at the top but is together with other 5 classifiers, which means that RNN has similar performance with them. Nevertheless, RNN has the best average $CE_{0.2}$ (0.534) over 9 datasets according to Table \ref{table.rq2a} and is 10\% more than the second technique J48 (0.484). The SK result together with average values supports the better performance of our approach compared with typical techniques.

Table \ref{table.rq2a} also lists the Win/Tie/Loss results under each $CE_\pi$. In most cases, RNN has more than half (5 out of 9) `Win' against other techniques. When compared with RF, RNN has better average value, but 4 `Win' and 4 `Loss' under $CE_{0.5}$ and $CE_{1.0}$, which means that RNN is not significantly better than RF in most datasets. This result indicates that RNN may not be suitable to use both code metrics and process metrics.

\subsubsection{RQ2b: How is the performance of RNN using only code metrics comparing with typical classifiers using both code metrics and process metrics?}
\label{section.rq2b}

Comparing the performance of RNN between Table \ref{table.rq1} and Table \ref{table.rq2a}, it is clear that RNN has better performance using HVSM built with only code metrics. Since process metrics are more difficult to achieve than code metrics, it is meaningful to compare the performance of RNN using only code metrics with typical classifiers using both code and process metrics. In this section, HVSM uses metric set in each version that consists of only the 20 code metrics, which is the same as the experiment in RQ1. 

According to the results in RQ2a, RF performs the best among the 7 typical techniques evaluated by each $CE_\pi$, so this paper selects RF on behalf of typical techniques to compare with RNN. Table \ref{table.rq2b} shows the detailed results of RNN (using only code metrics) and RF (using both code and process metrics) under each $CE_\pi$. From the table we can see that RNN has at least 6 (out of 9) better performance under different $CE_\pi$. When it comes to Win/Tie/Loss, RNN has 7 `Win' in $CE_{0.1}$ and $CE_{0.2}$ and 5 `Win' in $CE_{0.5}$ and $CE_{1.0}$, the number of `Loss' is no more than 3. This result shows that RNN (using only code metrics) outperforms RF (using both code and process metrics) with statistical significance in more than half of the datasets.

\textbf{Generally speaking, RNN has better performance compared with typical techniques using both code and process metrics. Furthermore, RNN using HVSM built with only code metrics outperforms baseline classifiers trained with both code and process metrics with statistical significance.}

\begin{table}[h]
	\centering
	\caption{The CE performance of RNN using HVSM built with only code metrics comparing with RF using both code and process metrics. Bold font highlights the better performance between RNN and RF.}
	\label{table.rq2b}%
	\resizebox{1\columnwidth}{!}{
		\begin{tabular}{l|cc|cc|cc|cc}
			\toprule
			& \multicolumn{2}{c|}{$CE_{0.1}$} & \multicolumn{2}{c|}{$CE_{0.2}$} & \multicolumn{2}{c|}{$CE_{0.5}$} & \multicolumn{2}{c}{$CE_{1.0}$} \\
			\midrule
			Target (Tr-$>$T) & RNN   & RF    & RNN   & RF    & RNN   & RF    & RNN   & RF \\
			\midrule
			ant 1.6-$>$1.7 & 0.023 & \textbf{0.056} & 0.049 & \textbf{0.072} & 0.125 & \textbf{0.14} & 0.209 & \textbf{0.221} \\
			camel 1.4-$>$1.6 & \textbf{0.354} & 0.302 & \textbf{0.396} & 0.334 & \textbf{0.462} & 0.418 & \textbf{0.533} & 0.48 \\
			jedit 4.2-$>$4.3 & \textbf{0.438} & 0.267 & \textbf{0.455} & 0.333 & \textbf{0.496} & 0.475 & 0.537 & \textbf{0.583} \\
			log4j 1.1-$>$1.2 & \textbf{0.757} & 0.612 & \textbf{0.715} & 0.558 & \textbf{0.706} & 0.537 & \textbf{0.719} & 0.516 \\
			lucene 2.2-$>$2.4 & \textbf{0.757} & 0.725 & \textbf{0.767} & 0.722 & \textbf{0.781} & 0.738 & \textbf{0.806} & 0.776 \\
			poi 2.5-$>$3.0 & \textbf{0.6} & 0.543 & \textbf{0.67} & 0.647 & \textbf{0.737} & 0.736 & \textbf{0.781} & 0.78 \\
			velocity 1.5-$>$1.6 & \textbf{0.538} & 0.508 & \textbf{0.571} & 0.542 & \textbf{0.694} & 0.679 & \textbf{0.764} & 0.753 \\
			xalan 2.5-$>$2.6 & 0.548 & \textbf{0.603} & 0.612 & \textbf{0.651} & 0.671 & \textbf{0.7} & 0.708 & \textbf{0.721} \\
			xerces 1.3-$>$1.4 & \textbf{0.701} & 0.495 & \textbf{0.764} & 0.498 & \textbf{0.824} & 0.439 & \textbf{0.848} & 0.23 \\
			\midrule
			Avg.  & \textbf{0.524 } & 0.457  & \textbf{0.555 } & 0.484  & \textbf{0.611 } & 0.540  & \textbf{0.656 } & 0.562  \\
			Win/Tie/Loss &  ---     & 7/0/2 &   ---    & 7/0/2 &  ---     & 5/2/2 &    ---   & 5/1/3 \\
			\bottomrule
		\end{tabular}%
	}
	
\end{table}%

\section{Discussion}
\label{section.discussion}
\subsection{Fairness of Training Data}
%Does RNN outperform other techniques with the help of more files involved in HVSM than that in single version used by other classifiers?

In our approach, RNN predicts defects in test set using HVSM as its training set. According to the definition of HVSM, it has access to the history of metrics in previous versions, while the training set of typical classifiers does not. The comparison between our approach and other techniques seems unfair. To be noticed that,  this paper is proposed to draw attention to using the historical information in previous versions (like HVSM) instead of using data in just one single version as typical techniques do in WPDP. The result in Section \ref{section.result} shows that our approach outperforms typical techniques using data in a single previous version in WPDP, and this section will show the result of typical techniques using data in the whole version sequence which is more fair comparing with our approach.

In this section, the training data of typical classifiers are mixed-up files in the whole version sequence. For example, when training data in \textbf{ant 1.6} and predicting defects in \textbf{ant 1.7}, the training set of typical classifiers consists of all files in \textbf{ant 1.3, 1.4, 1.5, 1.6}. Table \ref{table.discussion1} shows a clear review of the results comparing RNN with typical classifiers using mixed-up training data. 
%To be noticed that RNN here uses only code metrics (the same as RQ1). 
For training data of typical techniques, the metric set they use still has two types: (1) only code metrics (cm), (2) both code and process metrics (cm+pm). According to the result of RQ2b, RNN uses HVSM built with only code metrics in the following result. From the table we can see that with the whole version sequence included in the training data, typical classifiers still have worse average performance than RNN. This result supports the usefulness of sequential information that HVSM has, which is not included in the simple mixed-up training data used by typical techniques.

\begin{table}[htbp]
	\centering
	\caption{The average $CE_{\pi}$ value of RNN comparing with typical classifiers using mixed-up training data. The techniques are ranked by descending order of their average value of the 4 listed results}
	\label{table.discussion1}%
	\begin{threeparttable}
		\begin{tabular}{l|c|c|c|c}
			\toprule
		 	Technique & $CE_{0.1}$  & $CE_{0.2}$    & $CE_{0.5}$    & $CE_{1.0}$   \\
		 	\midrule
			RNN   & 0.524 \tnote{1}  & 0.555  & 0.611  & 0.656  \\
			\midrule
			LR(cm) & 0.515  & 0.545  & 0.592  & 0.629  \\
			LR(cm+pm) & 0.511  & 0.536  & 0.584  & 0.620  \\
			NN(cm) & 0.495  & 0.522  & 0.570  & 0.612  \\
			NN(cm+pm) & 0.480  & 0.503  & 0.545  & 0.587  \\
			RF(cm+pm) & 0.486  & 0.497  & 0.543  & 0.559  \\
			J48(cm) & 0.486  & 0.504  & 0.539  & 0.537  \\
			RF(cm) & 0.456  & 0.474  & 0.511  & 0.518  \\
			C5.0(cm+pm) & 0.414  & 0.436  & 0.464  & 0.428  \\
			C5.0(cm) & 0.396  & 0.412  & 0.430  & 0.378  \\
			J48(cm+pm) & 0.384  & 0.383  & 0.397  & 0.383  \\
			KNN(cm) & 0.163  & 0.131  & 0.041  & -0.195  \\
			KNN(cm+pm) & 0.151  & 0.132  & 0.051  & -0.198  \\
			NB(cm+pm) & 0.140  & 0.119  & 0.022  & -0.166  \\
			NB(cm) & 0.127  & 0.105  & 0.015  & -0.133  \\
			\bottomrule
		\end{tabular}%
		\begin{tablenotes}
			\footnotesize
			\item[1] The result is the average value of each technique's performance in the 9 datasets.
		\end{tablenotes}
	\end{threeparttable}
	
\end{table}%

\subsection{Performance Evaluated by Other Measures}

In addition to $CE_\pi$, this paper also provides performance of RNN and typical techniques evaluated by other 2 measures: AUC and ACC. According to the result of RQ2b, RNN uses HVSM built with code metrics only in the following results. 

\begin{figure}
	\centering
	\subfigure[AUC]{
		\label{figure.discussion2.auc} %% label for first subfigure
		\includegraphics[width=0.95\columnwidth]{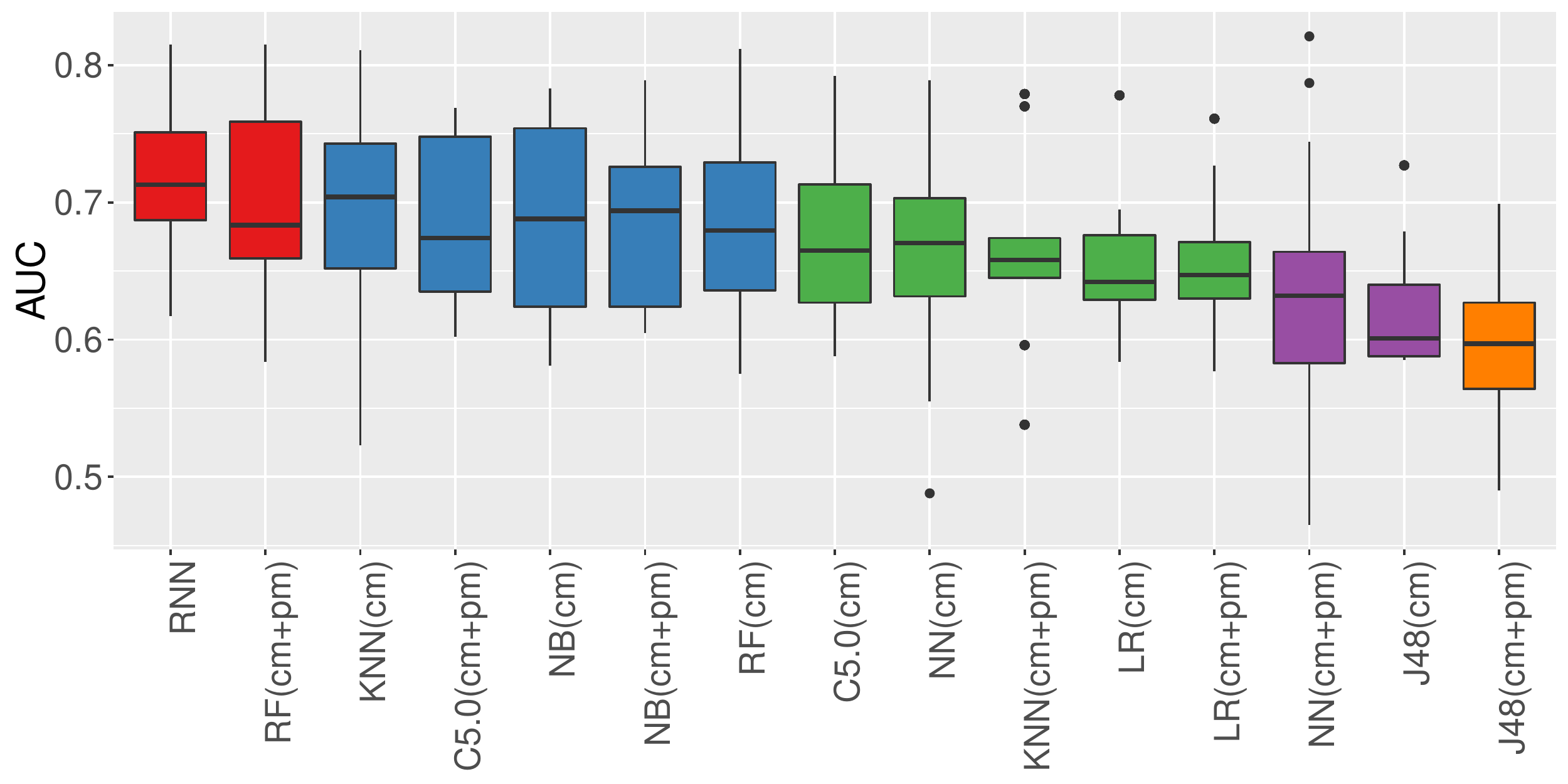}}
	\subfigure[ACC]{
		\label{figure.discussion2.acc} %% label for first subfigure
		\includegraphics[width=0.95\columnwidth]{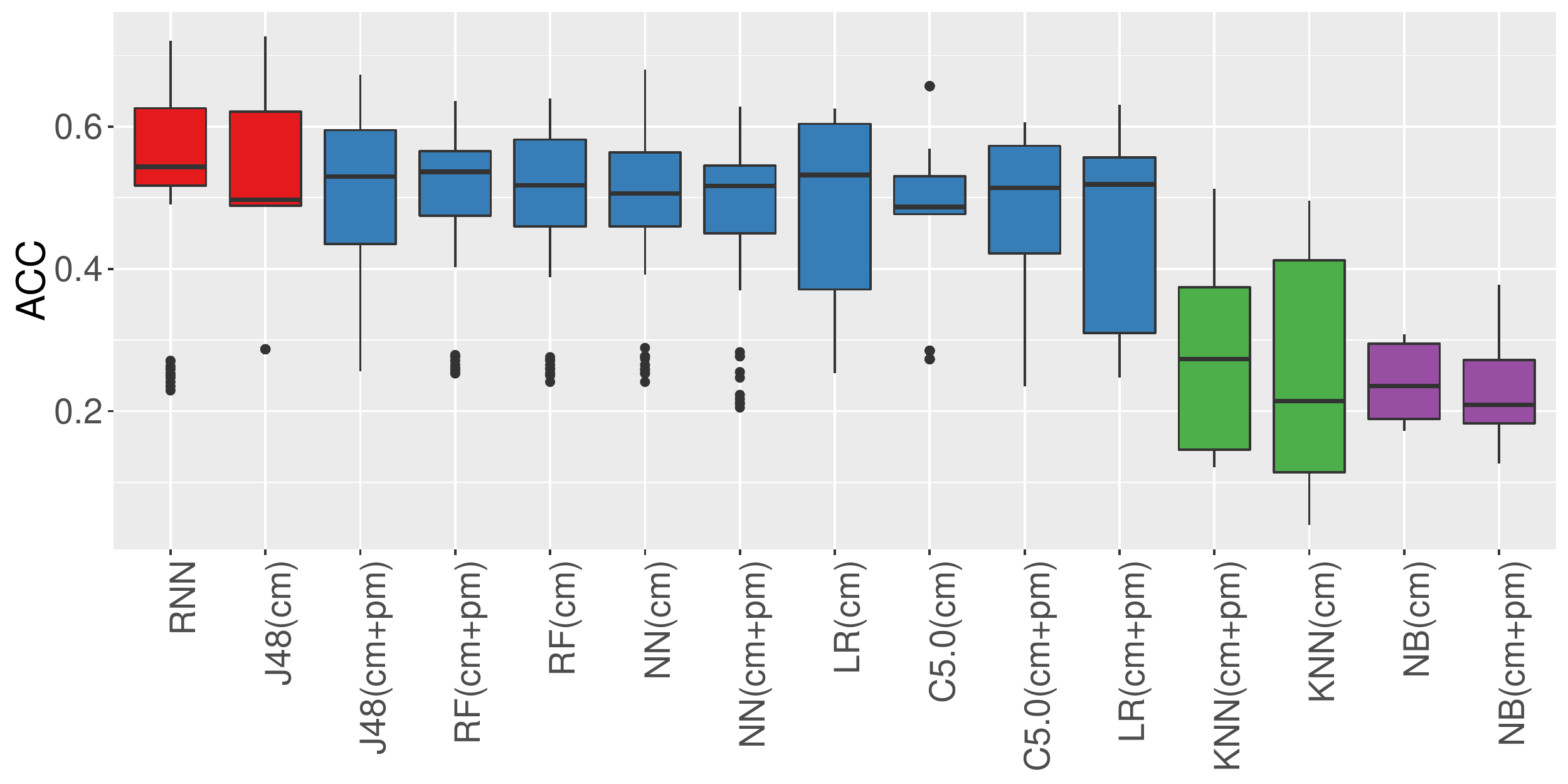}}
	\caption{The boxplots of  AUC and ACC values of 7 baseline classifiers using 2 types of metric sets(code metrics only(cm), code and process metrics(cm+pm)) and RNN. Different colors represents different Scott-Knott test ranks (from top down, the order is red, blue, green, purple, orange).}
	\label{figure.discussion2}
\end{figure}

\textbf{\emph{AUC}\hspace{1em}}
The \emph{Area Under the ROC Curve (AUC)} \cite{auc.origin} is calculated from the \emph{Receiver Operator Characteristic (ROC)} curve. AUC is a threshold-independent performance metric that plots the false positive rate ($\frac{FP}{FP+TN}$) against the true positive rate ($\frac{TP}{TP+FN}$). It measures how a classifier can discriminate between buggy and clean files, and the higher AUC value indicates a better performance. AUC is a widely used evaluation metric that was adopted by many works \cite{background.dp5,prediction.methodlevel,Lessmann.benchmark,background.dp3,related.CLAMI,background.wpdp,multiMetrics3,auc.use1,Rparameter}. Figure \ref{figure.discussion2.auc} shows the boxplots of performance of RNN and typical techniques under AUC. RNN is ranked at the top by SK test with the best average AUC over the tested datasets, and has distinct advantages compared with most of the typical classifiers.

\textbf{\emph{ACC}\hspace{1em}}
In addition to CE, ACC \cite{multiMetrics.YYB, Kamei2013} is another commonly used indicators that describe the effort-aware ranking effectiveness of a classification technique. ACC denotes the recall of defective files when using 20\% of the entire efforts according to its rank. Figure \ref{figure.discussion2.acc} shows the performance of RNN and typical techniques under ACC. It can be seen that RNN is at the top rank under SK test. This further supports the result in Section \ref{section.result}. 

In summary, evaluated by AUC and ACC, RNN still has better performance compared with other techniques. 
%especially when using HVSM built with code metrics only (i.e. RNN(cm) in the results). 

%Since F1 is a threshold-dependent performance metric, we apply a method introduced by  to choose a reasonable threshold. In practice, engineers balance between true positive rate ($TPR$) and false positive rate ($FPR$), and Menzies operationalizes this notion of balance:
%It is suggested that higher $bal$ means a better balance between $TPR$ and $FPR$  . In our work, we choose the F1 threshold that makes the $bal$ largest.

\section{Threats to Validity}
\label{section.threats}
\textit{\textbf{Project selection} } In this work, we select 9 open-source projects which have been used by prior works. The code metrics are extracted and validated by M. Jureczko \cite{Promise.data0}. The process metrics can be directly calculated from the source code of each project. Furthermore, it is recommended that more projects and metrics should be tested using our approach, and the result may vary.

\textit{\textbf{Techniques selection} } Most of the classification techniques that we use are commonly investigated in defect prediction literature. The study on more techniques is required. In addition, the classifier that we use to process HVSM is RNN, which is one of the techniques that can handle sequential data. Replication studies using different classifiers to process HVSM may prove fruitful.

\textit{\textbf{Study replication} } The typical techniques that we use as baselines are implemented using R packages. We will provide an open-source implementation of our RNN online. Besides, randomness in some classifiers including RNN will make replication a little different from our result. 

\section{Conclusions}
\label{section.conclusion}
Accurate software defect prediction could help software practitioners allocate test resources to defect-prone modules effectively and efficiently. In the last decades, much effort has been devoted to build accurate defect prediction models, including developing quality defect predictors and modeling techniques. However, current widely used defect predictors such as code metrics and process metrics could not well describe how software modules change over the project evolution, which we believe is important for defect prediction. In order to deal with this problem, we propose to use the Historical Version Sequence of Metrics (HVSM) in continuous software versions as defect predictors. Furthermore, we leverage Recurrent Neural Network (RNN), a popular modeling technique, to take HVSM as the input to build software prediction models. 
%The experimental results show that, in most cases, the proposed HVSM-based RNN model has a significantly higher prediction accuracy than the commonly used baseline models. In particular, the proposed HVSM-based RNN model can identify many real defects missed by the commonly used baseline models.
%Software defect predictions aims to detect fault-prone parts of the software and help to reduce the effort of testing and debugging. In order to improve the performance of defect prediction, a lot of attention has been attracted to find better classification metrics and techniques. In prior studies, researchers mainly focused on predicting defects using code metrics in a single version or process metrics across two versions as training data for classifiers. However, these two kinds of metrics could not well describe the historical trend that files change over the project's evolution. We hope to highlight the historical trend of files' changes expressed by the version sequence of projects. 

%In this paper, Historical Version Sequence of Metrics (HVSM) is proposed to describe the historical trend of file's changes across versions. It is a simple approach based on the code and process metrics that combines the metrics of the same file in ascending order of versions. We also provide a classification technique RNN to handle HVSM and perform defect prediction in CVDP. 

Our evaluation on 9 open source projects shows that our approach outperforms 7 baseline classifiers. We examine the results mainly in effort-aware scenarios measured by cost-effectiveness(CE). The Win/Tie/Loss evaluation with Wilcoxon signed-rank test and Cliff's delta $\delta$, and Scott-Knott test are also applied to support our results. In most cases, the proposed HVSM-based RNN model has a statistically significantly better effort-aware ranking effectiveness than the commonly used baseline models. In summary, our contributions are as follows:
\begin{itemize}
	\item \textbf{Providing HVSM to highlight the historical trend that files change in version sequence.} HVSM can describe a file's changing information in sequence by joining its metrics in a specific number of continuous historical versions.
	%From the statistics of our studied projects, HVSM can cover most of the file in a project's version history.
	% It is also convenient to extract HVSM from an existing dataset that contains static code metrics of file in each version of a project.
	\item \textbf{Leveraging a proper technique, RNN, to handle HVSM in defect prediction.} We apply RNN to HVSM to perform within-project defect prediction. The comparison between RNN and other baseline classifiers shows that our approach has better performance with statistical significance in effort-aware scenarios. In addition, it is suggested to use code metrics to build HVSM that RNN uses in order to achieve better performance.% We also encourage future works to apply different techniques to HVSM or using the information provided by HVSM to improve the performance of typical techniques.
\end{itemize}

In the future, we would like to extend our approach to more projects in defect prediction. In addition, we encourage future works to apply different techniques to HVSM or using the information provided by HVSM to improve the performance of typical techniques.

\bibliographystyle{abbrv}
\bibliography{defect_prediction}

\end{document}